\documentclass[12pt]{article}
\usepackage[dvips]{graphicx}
\usepackage{epsfig}
\usepackage{latexsym,amsmath,amsfonts,amssymb}
\usepackage[latin1]{inputenc}
\usepackage[american]{babel}
\usepackage[dvips]{graphicx}
\usepackage{bbm}
\usepackage{color}
\usepackage{slashed}
\usepackage[unicode]{hyperref}
\def\im{Invent. Math.}

\def\a{\alpha}
\def\b{\beta}
\def\c{\gamma}
\def\d{\delta}
\def\f{\phi}               
\def\vf{\varphi}  
\def\tvf{\tilde{\varphi}}
\def\vp{\varphi}
\def\g{\gamma}
\def\h{\eta}
\def\j{\psi}
\def\k{\kappa}                    
\def\l{\lambda}
\def\m{\mu}
\def\n{\nu}
\def\o{\omega}  \def\w{\omega}

\def\q{\theta}  \def\th{\theta}                  
\def\r{\rho}                                     
\def\s{\sigma}                                   
\def\t{\tau}
\def\u{\upsilon}
\def\x{\xi}
\def\z{\zeta}
\def\pt{\tilde{\varphi}}
\def\tt{\tilde{\theta}}
\def\lab{\label}
\def\6{\partial}
\def\wg{\wedge}
\def\bpsi{\bar{\psi}}
\def\bt{\bar{\theta}}
\def\bvf{\bar{\varphi}}

\DeclareMathOperator{\tr}{tr}

\newcommand{\be}{\begin{equation}}
\newcommand{\ee}{\end{equation}}
\newcommand{\beq}{\begin{equation}}
\newcommand{\eeq}{\end{equation}}
\newcommand{\bea}{\begin{eqnarray}}
\newcommand{\eea}{\end{eqnarray}}
\newcommand{\nn}{\nonumber}

\newcommand{\ba}{\begin{eqnarray}}
\newcommand{\ea}{\end{eqnarray}}

\newcommand{\beqs}{\begin{eqnarray}}
\newcommand{\eeqs}{\end{eqnarray}}
\newcommand{\bal}{\begin{aligned}}
\newcommand{\eal}{\end{aligned}}

\begin{document}
\baselineskip=15.5pt
\pagestyle{plain}
\setcounter{page}{1}

\def\del{{\partial}}
\def\vev#1{\left\langle #1 \right\rangle}
\def\cn{{\cal N}}
\def\co{{\cal O}}


\def\IC{{\mathbb C}}
\def\IR{{\mathbb R}}
\def\IZ{{\mathbb Z}}
\def\RP{{\bf RP}}
\def\CP{{\bf CP}}
\def\Poincare{{Poincar\'e }}
\def\tr{{\rm tr}}
\def\tp{{\tilde \Phi}}

\def\TL{\hfil$\displaystyle{##}$}
\def\TR{$\displaystyle{{}##}$\hfil}
\def\TC{\hfil$\displaystyle{##}$\hfil}
\def\TT{\hbox{##}}
\def\HLINE{\noalign{\vskip1\jot}\hline\noalign{\vskip1\jot}}
\def\seqalign#1#2{\vcenter{\openup1\jot
   \halign{\strut #1\cr #2 \cr}}}
\def\lbldef#1#2{\expandafter\gdef\csname #1\endcsname {#2}}
\def\eqn#1#2{\lbldef{#1}{(\ref{#1})}%
\begin{equation} #2 \label{#1} \end{equation}}
\def\eqalign#1{\vcenter{\openup1\jot
     \halign{\strut\span\TL & \span\TR\cr #1 \cr
    }}}

\def\eno#1{(\ref{#1})}
\def\href#1#2{#2}
\def\half{\frac{1}{2}}



\def\ads{{\it AdS}}
\def\adsp{{\it AdS}$_{p+2}$}
\def\cft{{\it CFT}}

\newcommand{\ber}{\begin{eqnarray}}
\newcommand{\eer}{\end{eqnarray}}

\newcommand{\beqar}{\begin{eqnarray}}
\newcommand{\cN}{{\cal N}}
\newcommand{\cO}{{\cal O}}
\newcommand{\cA}{{\cal A}}
\newcommand{\cT}{{\cal T}}
\newcommand{\cF}{{\cal F}}
\newcommand{\cC}{{\cal C}}
\newcommand{\cR}{{\cal R}}
\newcommand{\cW}{{\cal W}}
\newcommand{\eeqar}{\end{eqnarray}}
\newcommand{\tht}{\thteta}
\newcommand{\lm}{\lambda}\newcommand{\Lm}{\Lambda}


\newcommand{\nonu}{\nonumber}
\newcommand{\oh}{\displaystyle{\frac{1}{2}}}
\newcommand{\dsl}
   {\kern.06em\hbox{\raise.15ex\hbox{$/$}\kern-.56em\hbox{$\partial$}}}
\newcommand{\id}{i\!\!\not\!\partial}
\newcommand{\as}{\not\!\! A}
\newcommand{\ps}{\not\! p}
\newcommand{\ks}{\not\! k}
\newcommand{\D}{{\cal{D}}}
\newcommand{\dv}{d^2x}
\newcommand{\Z}{{\cal Z}}
\newcommand{\N}{{\cal N}}
\newcommand{\Dsl}{\not\!\! D}
\newcommand{\Bsl}{\not\!\! B}
\newcommand{\Psl}{\not\!\! P}

\newcommand{\eeqarr}{\end{eqnarray}}
\newcommand{\ZZ}{{\rm \kern 0.275em Z \kern -0.92em Z}\;}


\def\del{{\delta^{\hbox{\sevenrm B}}}} \def\ex{{\hbox{\rm e}}}
\def\azb{A_{\bar z}} \def\az{A_z} \def\bzb{B_{\bar z}} \def\bz{B_z}
\def\czb{C_{\bar z}} \def\cz{C_z} \def\dzb{D_{\bar z}} \def\dz{D_z}
\def\im{{\hbox{\rm Im}}} \def\mod{{\hbox{\rm mod}}} \def\tr{{\hbox{\rm Tr}}}
\def\ch{{\hbox{\rm ch}}} \def\imp{{\hbox{\sevenrm Im}}}
\def\trp{{\hbox{\sevenrm Tr}}} \def\vol{{\hbox{\rm Vol}}}
\def\rl{\Lambda_{\hbox{\sevenrm R}}} \def\wl{\Lambda_{\hbox{\sevenrm W}}}
\def\fc{{\cal F}_{k+\cox}} \def\vev{vacuum expectation value}
\def\nodiv{\mid{\hbox{\hskip-7.8pt/}}}
\def\ie{{\em i.e.}}
\def\ie{\hbox{\it i.e.}}

\def\CC{{\mathchoice
{\rm C\mkern-8mu\vrule height1.45ex depth-.05ex
width.05em\mkern9mu\kern-.05em}
{\rm C\mkern-8mu\vrule height1.45ex depth-.05ex
width.05em\mkern9mu\kern-.05em}
{\rm C\mkern-8mu\vrule height1ex depth-.07ex
width.035em\mkern9mu\kern-.035em}
{\rm C\mkern-8mu\vrule height.65ex depth-.1ex
width.025em\mkern8mu\kern-.025em}}}

\def\RR{{\rm I\kern-1.6pt {\rm R}}}
\def\NN{{\rm I\!N}}
\def\ZZ{{\rm Z}\kern-3.8pt {\rm Z} \kern2pt}
\def\IB{\relax{\rm I\kern-.18em B}}
\def\ID{\relax{\rm I\kern-.18em D}}
\def\II{\relax{\rm I\kern-.18em I}}
\def\IP{\relax{\rm I\kern-.18em P}}
\newcommand{\CS}{{\scriptstyle {\rm CS}}}
\newcommand{\CSs}{{\scriptscriptstyle {\rm CS}}}
\newcommand{\rc}{\nonumber\\}
\newcommand{\bear}{\begin{eqnarray}}
\newcommand{\eear}{\end{eqnarray}}

\newcommand{\LL}{{\cal L}}

\def\mani{{\cal M}}
\def\calo{{\cal O}}
\def\calb{{\cal B}}
\def\calw{{\cal W}}
\def\calz{{\cal Z}}
\def\cald{{\cal D}}
\def\calc{{\cal C}}

\def\to{\rightarrow}
\def\ele{{\hbox{\sevenrm L}}}
\def\ere{{\hbox{\sevenrm R}}}
\def\zb{{\bar z}}
\def\wb{{\bar w}}
\def\nodiv{\mid{\hbox{\hskip-7.8pt/}}}
\def\menos{\hbox{\hskip-2.9pt}}
\def\dr{\dot R_}
\def\drr{\dot r_}
\def\ds{\dot s_}
\def\da{\dot A_}
\def\dga{\dot \gamma_}
\def\ga{\gamma_}
\def\dal{\dot\alpha_}
\def\al{\alpha_}
\def\cl{{closed}}
\def\cls{{closing}}
\def\vev{vacuum expectation value}
\def\tr{{\rm Tr}}
\def\to{\rightarrow}
\def\too{\longrightarrow}


\def\a{\alpha}
\def\b{\beta}
\def\c{\gamma}
\def\d{\delta}
\def\e{\epsilon}           
\def\F{\Phi}
\def\f{\phi}               
\def\vf{\varphi}  \def\tvf{\tilde{\varphi}}
\def\vp{\varphi}
\def\g{\gamma}
\def\h{\eta}
\def\j{\psi}
\def\k{\kappa}                    
\def\l{\lambda}
\def\m{\mu}
\def\n{\nu}
\def\o{\omega}  \def\w{\omega}
\def\q{\theta}  \def\th{\theta}                  
\def\r{\rho}                                     
\def\s{\sigma}                                   
\def\t{\tau}
\def\u{\upsilon}
\def\x{\xi}
\def\X{\Xi}
\def\z{\zeta}
\def\pt{\tilde{\varphi}}
\def\tt{\tilde{\theta}}
\def\lab{\label}
\def\6{\partial}
\def\wg{\wedge}
\def\atanh{{\rm arctanh}}
\def\bpsi{\bar{\psi}}
\def\bt{\bar{\theta}}
\def\bvf{\bar{\varphi}}

%



\newfont{\namefont}{cmr10}
\newfont{\addfont}{cmti7 scaled 1440}
\newfont{\boldmathfont}{cmbx10}
\newfont{\headfontb}{cmbx10 scaled 1728}





\newcommand{\re}{\,\mathbb{R}\mbox{e}\,}
\newcommand{\hyph}[1]{$#1$\nobreakdash-\hspace{0pt}}
\providecommand{\abs}[1]{\lvert#1\rvert}
\newcommand{\Nugual}[1]{$\mathcal{N}= #1 $}
\newcommand{\sub}[2]{#1_\text{#2}}
\newcommand{\partfrac}[2]{\frac{\partial #1}{\partial #2}}
\newcommand{\bsp}[1]{\begin{equation} \begin{split} #1 \end{split} \end{equation}}
\newcommand{\calF}{\mathcal{F}}
\newcommand{\calO}{\mathcal{O}}
\newcommand{\calM}{\mathcal{M}}
\newcommand{\calV}{\mathcal{V}}
\newcommand{\bbZ}{\mathbb{Z}}
\newcommand{\bbC}{\mathbb{C}}
\newcommand{\cK}{{\cal K}}

\newcommand{\Thq}{\Theta\left(\r-\r_q\right)}
\newcommand{\Dq}{\d\left(\r-\r_q\right)}
\newcommand{\kten}{\kappa^2_{\left(10\right)}}
\newcommand{\pbi}[1]{\imath^*\left(#1\right)}
\newcommand{\ho}{\hat{\omega}}
\newcommand{\tth}{\tilde{\th}}
\newcommand{\tf}{\tilde{\f}}
\newcommand{\tj}{\tilde{\j}}
\newcommand{\tw}{\tilde{\omega}}
\newcommand{\tz}{\tilde{z}}
\newcommand{\prj}[2]{(\partial_r{#1})(\partial_{\j}{#2})-(\partial_r{#2})(\partial_{\j}{#1})}
\def\atanh{{\rm arctanh}}
\def\sech{{\rm sech}}
\def\csch{{\rm csch}}
\allowdisplaybreaks[1]

\def\red{\textcolor[rgb]{0.98,0.00,0.00}}

\newcommand{\Dan}[1] {{\textcolor{blue}{#1}}}

\numberwithin{equation}{section}

\newcommand{\Tr}{\mbox{Tr}}    


%

\renewcommand{\theequation}{{\rm\thesection.\arabic{equation}}}

\setcounter{footnote}{0}
\renewcommand{\theequation}{{\rm\thesection.\arabic{equation}}}

\begin{titlepage}

\hfill FPAUO-15/06  \\

\phantom{xx}
\vskip 0.4in

\begin{center}
{\Large \bf  {A $\mathcal{N}=2$ Supersymmetric $AdS_4$ Solution in M-theory }}
\vskip 0.1in

{\Large \bf  {with Purely Magnetic Flux}}

\vskip 0.4in

{\bf Yolanda Lozano}${}^{a,}{}^{1}$,~{\bf Niall T. Macpherson}${}^{b,}{}^{2}$,~{\bf Jes\'us Montero}${}^{a,}{}^{3}$
\\

\vskip .2in

a: Department of Physics,  University of Oviedo,\\
Avda.~Calvo Sotelo 18, 33007 Oviedo, Spain\\

\vskip .2in

b: Dipartimento di Fisica, Universit\`a di Milano--Bicocca, I-20126 Milano, Italy\\
    and\\
    INFN, sezione di Milano--Bicocca\\
 
\vskip 5mm

\vspace{0.2in}
\end{center}
\vspace{0.2in}
\centerline{{\bf Abstract}}
We find a new $\mathcal{N}=2$ $AdS_4$ solution in M-theory supported by purely magnetic flux via a sequence of abelian and non-abelian T-dualities. This provides the second known example in this class besides the uplift of the Pernici and Sezgin solution to 7d gauged supergravity constructed in the eighties. We compute the free energy of the solution, and show that it scales as $N^{3/2}$. It is intriguing that even though the natural holographic interpretation is in terms of M5-branes wrapped on a special Lagrangian 3-cycle,  this solution does not exhibit the expected $N^3$ behavior.
\smallskip

\smallskip

\vfill
\noindent
 {
 $^1$ylozano@uniovi.es,
$^2$niall.macpherson@mib.infn.it,
$^3$monteroaragon@uniovi.es}

\end{titlepage}
\setcounter{footnote}{0}

\tableofcontents

\setcounter{footnote}{0}
\renewcommand{\theequation}{{\rm\thesection.\arabic{equation}}}

\section{Introduction}

In recent years non-Abelian T duality (NAT duality) has been very successfully applied as a generator of new supergravity backgrounds that may have interesting applications in the context of the AdS/CFT correspondence \cite{Sfetsos:2010uq}-\cite{Bea:2015fja}. While some of these backgrounds represent explicit new solutions to existing classifications \cite{Sfetsos:2010uq, Itsios:2012zv, Itsios:2013wd, Sfetsos:2014tza}, some of them have been shown to fall outside known classifications \cite{Lozano:2015bra} or  to provide the only explicit solution to some set of PDEs \cite{Lozano:2012au}.

A very inspiring example is the $AdS_6$ solution to Type IIB supergravity constructed in \cite{Lozano:2012au}. Supersymmetry is known to impose strong constraints on $AdS_6$ backgrounds \cite{Passias:2012vp, Apruzzi:2014qva}\footnote{See also \cite{Kim:2015hya}.}, even if large classes of fixed point theories are known to exist in 5 dimensions \cite{Seiberg:1996bd, Morrison:1996xf, 
Intriligator:1997pq} with expected $AdS_6$ duals. The only $AdS_6/CFT_5$ explicit example identified to date is the duality between the Brandhuber and Oz solution to massive Type IIA \cite{Brandhuber:1999np} (known to be the only possible IIA $AdS_6$ background \cite{Passias:2012vp}\footnote{Variations of it such as orbifold solutions have also been constructed in \cite{Bergman:2012kr}.}) and the fixed point theory that arises from the D4/D8/O8 system in \cite{Seiberg:1996bd}. Yet, there are families of 5 and 7-brane webs giving rise to 5d fixed point theories   \cite{Aharony:1997ju, Aharony:1997bh, DeWolfe:1999hj} whose dual $AdS_6$ spaces remain to be identified. The solution in \cite{Lozano:2012au}\footnote{See \cite{Lozano:2013oma} for  a discussion of the properties of the associated CFT.} provides a possible holographic dual to these theories.

The duality between 3d SCFTs arising from M5-branes wrapped on 3d manifolds and $AdS_4$ spaces is yet another example in which explicit $AdS_4$ solutions with the required properties are scarce. Remarkable progress has been achieved recently \cite{Rota:2015aoa} through the construction of explicit $AdS_4 \times \Sigma_3 \times M_4$ solutions to massive IIA which are candidate duals to compactifications of the $(1,0)$ 6d CFTs living in NS5-D6-D8 systems \cite{Gaiotto:2014lca} on a 3-manifold $\Sigma_3$, which could eventually lead to a generalization of the 3d-3d correspondence  \cite{Terashima:2011qi} to $\mathcal{N}=1$. The $\mathcal{N}=2$ case is yet especially interesting, since with this number of supersymmetries the 3d-3d correspondence allows to associate a 3d $\mathcal{N}=2$ SCFT to the 3d manifold on which the M5-branes are wrapped \cite{Terashima:2011qi}. 
 This field theory arises as a twisted compactification on the 3d Riemann surface of the $(2,0)$ ${\rm CFT}_6$ living in the M5-branes. 

However, to date only one $\mathcal{N}=2$ $AdS_4$ explicit solution to M-theory is known that could provide the holographic dual to these compactifications. This solution is the uplift to eleven dimensions \cite{Acharya:2000mu, Gauntlett:2000ng}
of the Pernici-Sezgin solution \cite{Pernici:1984nw} to 7d gauged supergravity, that dates back to the 80's. This is of the form $AdS_4 \times M_7$ where $M_7$ is an $S^4$-fibration over a hyperbolic manifold $H_3$, on which the M5-branes are wrapped. The Pernici-Sezgin solution is the only explicit solution of the form $AdS_4 \times \Sigma_3\times S^4$ in the general class of $\mathcal{N}=2$ $AdS_4$ backgrounds obtained from M5-branes wrapping calibrated cycles in  \cite{Gauntlett:2006ux}. 

In this paper we construct a new $\mathcal{N}=2$ $AdS_4$ solution to M-theory belonging to the
general class of $\mathcal{N}=2$ $AdS_4$ backgrounds derived in  \cite{Gauntlett:2006ux}. This class is defined by requiring that the Killing spinors satisfy the same projection conditions as the wrapped branes and that there is no electric flux. Yet the solutions need not describe in general M5-branes wrapped in 3d manifolds in the near horizon limit. Our solution seems to belong to this more general class. 

We obtain our solution through non-Abelian T-duality on the $AdS_4\times CP^3$ background dual to ABJM  \cite{Aharony:2008ug},  followed by an Abelian T-duality and an uplift to eleven dimensions. The non-Abelian T-duality transformation is responsible for the breaking of the supersymmetries from $\mathcal{N}=6$ to $\mathcal{N}=2$. The detailed properties of the resulting $\mathcal{N}=2$ $AdS_4$ solution to Type IIB were studied in \cite{Lozano:2014ata}. This solution contains two $U(1)'s$, one of which can be further used to (Abelian) T-dualize back to Type IIA without breaking any of the supersymmetries. Finally, the solution is uplifted to eleven dimensions, where it can be shown to fulfill the conditions for 11d $\mathcal{N}=2$ $AdS_4$ solutions with purely magnetic flux, derived in \cite{Gauntlett:2006ux}\footnote{A systematic study of the most general class of $\mathcal{N}=2$ $AdS_4$ solutions of 11d supergravity, that includes the results in \cite{Gauntlett:2006ux}, was carried out in \cite{Gabella:2012rc}.}.

The paper is organized as follows. In section 2 we recall briefly the IIB solution constructed in \cite{Lozano:2014ata} through non-Abelian T-duality acting on the $AdS_4\times CP^3$ IIA background. In section 3 we construct its IIA Abelian T-dual, and discuss some properties of the associated dual CFT of relevance for the CFT interpretation of the 11d solution.
Section 4 contains the uplift to M-theory. 
Here we discuss some properties of the CFT associated to the 11d solution, that are implied by the analysis of the supergravity solution as well as its IIA description. 
We compute the holographic central charge and show that, as expected, it coincides with the central charge of the IIB solution written in terms of the 11d charges. Thus, it scales with $N^{3/2}$, contrary to the expectation for M5-branes.  We argue that the field theory analysis that we perform suggests that there should be Kaluza-Klein monopoles sourcing the background, and that M5-branes should only play a role in the presence of large gauge transformations (in a precise way that we define). This is intimately related to the existence of a non-compact direction inherited by the NAT duality transformation, which, as discussed at length in the NAT duality literature (see for instance  \cite{Lozano:2013oma, Lozano:2014ata, Bea:2015fja}), represents the most puzzling obstacle towards a precise CFT interpretation of this transformation. 
Finally, in section 6 we present our conclusions. Here  we discuss further our result for the free energy, as well as the view that we have taken to try to give a CFT meaning to the non-compact direction. We have relegated most of the technical details to three appendices. In Appendix A we include some details of the derivation of both the NAT and T dual solutions presented in sections 2 and 3. These details are especially relevant for the supersymmetry analysis. In Appendix B we review the G-structure conditions for preservation of supersymmetry of $AdS_4 \times M_6$ solutions to Type II supergravities. In Appendix C we perform the detailed supersymmetry analysis of the solutions in IIA, IIB and M-theory.

\section{The IIB NAT dual $AdS_4$ solution}

This solution was constructed in \cite{Lozano:2014ata}, where some properties of the associated dual CFT were also analyzed. We refer the reader to this paper for more details. In this section we present the background for completeness. More technical properties of the derivation that will be useful for the study of the backgrounds constructed from this one in the following sections are presented in Appendix A.

The background arises as the NAT dual of the $AdS_4\times \mathbb{CP}^3$ background with respect to a freely acting $SU(2)$ in the parameterization of the $\mathbb{CP}^3$ as a foliation in $T^{1,1}=S^2\times S^3$:
\beq\label{eq: metricbefore2} 
\begin{array}{l l}
ds^2 (\mathbb{CP}^3)&=d\zeta^2+\frac{1}{4}\bigg(\cos^2\zeta(d\theta_1^2+\sin^2\theta_1 d\phi_1^2)+\sin^2\zeta(d\theta_2^2+\sin^2\theta_2 d\phi_2^2)+\\
&~~~~~~~~~~~~~~~~~~~~~~~~~~~~~~~~~~~~+\sin^2\zeta\cos^2\zeta(d\psi+\cos\theta_1 d\phi_1+\cos\theta_2 d\phi_2)^2\bigg)\\
&=d\zeta^2+\frac{1}{4}\bigg(\cos^2\zeta(d\theta_1^2+\sin^2\theta_1 d\phi_1^2)+\sin^2{\zeta}(\o_1^2+\o_2^2)+\sin^2{\zeta}\cos^2{\zeta}(\o_3+\cos\theta_1 d\phi_1)^2\bigg)
\end{array}
\eeq
where $0\leq \zeta < \frac{\pi}{2},~0\leq \theta_i<\pi,~0\leq\phi_i\leq2\pi,~0\leq \psi \leq 4\pi$.

Dualising with respect to the $SU(2)$ acting on the 3-sphere parameterized by $(\psi,\theta_2,\phi_2)$ we obtain
\begin{equation}
\label{metric1}
d{\tilde s}^2=\frac{L^2}{4} ds^2 (AdS_4)+L^2\Bigl(d\zeta^2+\frac14 \cos^2{\zeta}\,(d\theta_1^2+\sin^2{\theta_1}d\phi_1^2)\Bigr)+ds^2 (M_3)\, ,
\end{equation}
where $ds^2 (M_3)$ stands for the 3-dimensional metric: 
\begin{eqnarray}
\label{dualmetric}
ds^2 (M_3)&=&\frac{1}{16\,  \det M}\Bigl[
L^4  \sin^4{\zeta} \Bigl(dr^2+r^2 d\chi^2-\sin^2{\zeta}\, (\sin{\chi}dr+r\cos{\chi}d\chi)^2+\nonumber\\
&&+r^2\cos^2{\zeta}\sin^2{\chi}\, (d\xi+\cos{\theta_1}d\phi_1)^2\Bigr)+  16 r^2dr^2\Bigr]\, .
\end{eqnarray}
${\rm det}M$ is given by:
\begin{equation}
 \det M=\frac{L^2}{64} \sin^2{\zeta} \Bigl(16  r^2(\sin^2{\chi}+ \cos^2{\chi}\cos^2{\zeta})+ L^4  \sin^4{\zeta}\cos^2{\zeta}\Bigr)\, .
\end{equation}
Here $(\chi,\xi)$ parameterize the new 2-sphere arising through the NAT duality transformation, that we will denote by ${\tilde S}^2$. $r$ is the non-compact coordinate generated by the transformation, which lives in $\mathbb{R}^+$. The presence of this non-compact direction is intimately related to the long-standing open problem of extending NAT duality beyond spherical world sheets. In the context of AdS/CFT applications this poses a problem to the CFT interpretation of AdS backgrounds generated through this transformation. Some ideas to provide a consistent interpretation have been proposed in \cite{Lozano:2013oma,Lozano:2014ata} (see also \cite{Bea:2015fja}), which we will partially  use in this paper. The reader is referred to these papers for more details.

The dilaton reads in turn
\begin{equation}
\label{dilaton}
e^\phi=\frac{L}{k}\frac{1}{\sqrt{{\rm det}M}}\, .
\end{equation}
A $B_2$ field is also generated that reads:
\begin{eqnarray}
&&B_2=\frac{L^2  \sin^2{\zeta}}{64\, {\rm det}M}\Bigl[ -L^4 r \cos^2{\zeta}\sin^4{\zeta}\cos{\theta_1}\sin{\chi}\, d\phi_1\wedge d\chi-\nonumber\\
&&-16\,  r^2 \, \Bigl( r (\cos^2{\zeta}\cos^2{\chi}+\sin^2{\chi})\, {\rm Vol}({\tilde S}^2)+\sin^2{\zeta}\sin^2{\chi}\cos{\chi}\, d\xi\wedge dr\Bigr) - \nonumber\\
&&-\cos^2{\zeta}\cos{\theta_1}\cos{\chi}\,(L^4  \sin^4{\zeta}+16  r^2)\,  dr\wedge d\phi_1\Bigr]\, .
\end{eqnarray}

Together with this we find the RR sector:
\begin{equation}
\label{RR1}
F_1=\frac{k}{2}   \Bigl(r   \sin^2{\zeta} \sin{\chi}\,  d\chi - \sin^2{\zeta} \cos{\chi}\, dr- r   \sin {2\zeta}  \cos{\chi}\, d{\zeta}\Bigr)
\end{equation}
\begin{eqnarray}
\label{RR2}
{\hat F}_3&=&-\frac{3}{128} k   L^4  \sin^3{2\zeta}  \,d\zeta\wedge 
{\rm Vol}(S^2_1)
+\frac{k}{2}  \Bigl(r \, dr \wedge ( \cos ^2{\zeta} \,  {\rm Vol}(S^2_1)+ \\
&&+ \sin{2\zeta}  \cos{\theta_1}\, d \zeta  \wedge d \phi_1+  \sin{2\zeta}  \sin ^2{\chi} \,d\zeta \wedge d\xi)- r^2 \sin{2\zeta}  \cos{\chi}\, d\zeta \wedge {\rm Vol} ({\tilde S}^2)  \Bigr) \nonumber
\end{eqnarray}
\begin{eqnarray}
\label{RR3}
{\hat F}_5&=& \frac{1}{64} k L^6   \sin^3{\zeta} \cos{\zeta} \, {\rm Vol}(AdS_4) \wedge d\zeta   -\frac{3}{8} k  L^2   r  \, {\rm Vol}(AdS_4)\wedge dr+\nonumber\\
&&+\frac{k}{2}     r^2  \Bigl(\cos ^2{\zeta}\, {\rm Vol}(S^2_1)
 +\sin{2\zeta}  \cos{\theta_1}  \, d\zeta  \wedge d \phi_1\Bigr) \wedge dr \wedge {\rm Vol}({\tilde S}^2)\, ,
\end{eqnarray}
where $F_p=d\, C_{p-1}-H_3\wedge C_{p-3}$ and ${\hat F}=F \wedge e^{-B_2}$  are the fluxes associated to the Page charges. 

Note that after the dualisation a singularity has appeared at the fixed point $\zeta=0$, where the squashed $S^3$ used to dualise shrinks to zero size. This singularity is associated to the component of the metric on the $r$-direction, and is always compensated with the singularity in the dilaton in  computations of physical quantities such as gauge couplings, internal volumes, etc. We will see that it will be inherited by the IIA and M-theory solutions where physical quantities will however be perfectly well defined as well.

\section{The IIA NAT-T Dual $AdS_4$ solution}
\label{sec: newIIASolution}

Following the steps in Appendix A we get the following solution in Type IIA after dualizing the previous background along the 
$\phi_1$ direction, that we will simply rename as $\phi$ \footnote{Also $\theta_1\equiv \theta$ and $S^2\equiv {\tilde S}^2$.}
\beq
\label{IIAmetric}
ds^2= \frac{L^2}{4} ds^2(AdS_4)+ L^2 d\zeta^2+ \frac{L^2}{4}\cos^2\zeta d\theta^2+ \sum_{i=1}^4 \left(\mathcal{G}^i\right)^2,
\eeq
where
\begin{align}\label{eq: Gs}
\mathcal{G}^1=&\frac{L}{2\sqrt{\Xi}}y_1 \sin^2\zeta \cos\zeta \sin\theta d\xi ,\nn\\[2mm]
\mathcal{G}^2=&-\frac{2}{ L \sqrt{\Delta}\sqrt{\mathcal{Z}}}\big(\mathcal{Z}dy_1+y_1 y_2 dy_2\big),\nn\\[2mm]
\mathcal{G}^3=&-\frac{L}{2\sqrt{\mathcal{Z}}} \sin\zeta dy_2,\nn\\[2mm]
\mathcal{G}^4=&\frac{2}{L\cos\zeta \sqrt{\Delta}\sqrt{\Xi}}\bigg[\Delta d\phi - \sin^2\zeta\cos^2\zeta \cos\theta   \left\{ y_1 y_2 dy_1 + \left(y_2^2+ \frac{L^4}{16}\sin^4\zeta\right)dy_2 \right\}\bigg]\, ,
\end{align}
and we have defined
\begin{align}
 \Delta&= \sin^2\zeta\big(y_1^2+\cos^2\zeta y_2^2 +\frac{L^4}{16}\sin^4\zeta\cos^2\zeta\big),~~~\Xi= \Delta\sin^2\theta + y_1^2 \sin^4\zeta\cos^2\theta, \\
 \mathcal{Z}&= y_1^2+ \frac{L^4}{16} \sin^4\zeta\cos^2\zeta, \nn
\end{align}
and
\beq
y_1=r \sin\chi,~~~ y_2 = r \cos\chi\, ,
\eeq
so that we have
\begin{equation}
 \begin{split}
  \label{Gis}
 64 L^2\sum_{i=1}^4 \left(\mathcal{G}^i\right)^2 
 & =  \frac{1}{\Delta\,\Xi}\cos^2\zeta  \Big\{ 16\Delta \sec^2\zeta d\phi  +  \sin^2\zeta \cos\theta \left[ L^4 \sin^4\zeta (\cos\chi dr - r \sin\chi d\chi) \right.  \\ 
 & \left.  + 16r^2 \cos\chi dr \right] \Big\}^2   
  + \frac{16 L^4}{\Xi} r^2 \sin^4\zeta \cos^2\zeta \sin^2\theta \sin^2\chi \,d\xi^2   \\ 
 & +  \frac{16 L^4}{\mathcal{Z}} \sin ^2\zeta (\cos\chi \,dr - r\sin\chi\, d\chi)^2 \\ 
 & +  \frac{256}{\Delta \, \mathcal{Z}} \left[ \frac{L^4}{16} \sin^4\zeta \cos^2\zeta (\sin\chi dr + r \cos\chi d\chi)  + r^2 \sin\chi dr \right]^2 \;.
 \end{split} 
\end{equation}

The NS 2 form is given by
\begin{align}\label{eq:B2_NATD_T}
B_2=&\frac{r^2}{\Xi}\sin^2\zeta \sin\chi\bigg[\sin^2\zeta \sin\chi\big(\cos\theta d\xi\wedge d\phi-\cos\chi d\xi\wedge dr\big)\nn\\
&-r\big(\cos^2\zeta\sin^2\theta +\sin^2 \zeta\sin^2\chi\big)d\chi\wedge d\xi\bigg]\;,
\end{align}
while the dilaton is
\beq
e^{\Phi}=\frac{4}{k L\cos\zeta\sqrt{\Xi}}\, .
\eeq
Notice that this blows up at $\zeta=0$ indicating that the geometry is singular here, which is confirmed when one studies the curvature invariants.

The RR sector is given by
\begin{align}
\hat{F}_2=&\frac{k}{16}\bigg[3L^4 \sin^3\zeta\cos^3\zeta\sin\theta d\zeta\wedge d\theta+ 8 r\sin2\zeta\big(\cos\theta d\zeta\wedge d r- \cos\chi d\zeta \wedge d \phi\big)\nn\\
&+8 r \cos^2\zeta\sin\theta d\theta\wedge d r - 8\sin^2\zeta\big(\cos\chi d r \wedge d \phi + r \sin\chi d\phi \wedge d \chi\big),\\[2mm]
\hat{F}_4=&\frac{k}{2}r \cos\zeta \sin\chi\bigg[2\sin\zeta\bigg(\sin\chi d\zeta\wedge d\xi\wedge dr \wedge d \phi+r \cos\theta d\zeta\wedge d\xi \wedge dr\wedge d\chi \nn\\
&\label{F2F4} r \cos\chi d\zeta \wedge d \xi \wedge d \phi \wedge d\chi\bigg)+r \cos\zeta \sin\theta d\theta \wedge d \xi \wedge dr \wedge d\chi\bigg],
\end{align}
where the gauge invariant fluxes are expressed in terms of these via $\hat{F}= F\wedge e^{-B_2}$.

\subsection{Supersymmetry}

It was shown in \cite{Lozano:2014ata,Kelekci:2014ima} that the NAT dual of ABJM preserves $\mathcal{N}=2$ supersymmetry in 3d, which means the R-symmetry is $U(1)$ in the dual CFT. The argument relies on a proof from \cite{Kelekci:2014ima}. In order to see this we must package all the dependence of the original geometry on the $SU(2)$ isometry in a  canonical frame\footnote{We write $e^{a+3}$ to match notation elsewhere where the canonical vielbeins are $e^4,e^5,e^6$.}
\beq
e^{a+3}=e^{C_a(x)}(\omega_a+A_a(x))
\eeq
where each left-invariant one form $\omega_a$ appears only once, and $x^{\mu}$ are coordinates on some  7d base which fibers the squashed 3-sphere containing the $SU(2)$ isometry. Then there is a bijective map between spinors independent of the $SU(2)$ directions in this frame and those preserved by the NAT dual solution. The map acts on the 10 dimensional MW Killing spinors as
\beq
\hat\epsilon_1 = \epsilon_1,~~~\hat\epsilon_2= \Omega_{SU(2)}\epsilon_2,
\eeq
with the matrix\footnote{This expression originally appeared in \cite{Itsios:2013wd}, where it was conjectured to hold by analogy with the Abelian case.}
\beq
\Omega_{SU(2)} = \Gamma^{(10)}\frac{-e^{C_1+C_2+C_3}\Gamma^{456}+v_ae^{C_a} \Gamma^{a+3}}{\sqrt{e^{2(C_1+C_2+C_3)}+ e^{2C_a} v_a^2}},
\eeq
where $v_a$ are dual coordinates in the NAT dual geometry, which we are expressing elsewhere in terms of spherical or cylindrical polar coordinates $v_1= y_1\cos\xi,~v_2= y_2\sin\xi,~v_3=y_2$.

In Appendix \ref{sec:SUSYdetailsABJM} we derive a spinor for ABJM independent of the $SU(2)$ directions. This may be written in terms of 6 dimensional MW spinors on $\mathbb{CP}^3$ as in Appendix \ref{sec: G-stucutureConvensions}
\beq
\eta^1_+= e^{i\frac{3\pi}{4}}\eta_+,~~~\eta^2_+=e^{-i\frac{3\pi}{4}}\eta_+
\eeq
where $(\eta^{1,2}_+)^* = \eta^{1,2}_-$ with the sign labeling chirality.  It is possible to decompose the 6d spinors in terms of two in linearly independent parts $\eta_+ = \pi_++\tilde{\pi}_+$ obeying the projections of eq (\ref{eq:piprojections}). We can then make the coordinate dependence explicit as
\beq\label{eq: dependence}
 \pi_+ = e^{\zeta \gamma^{34}}\pi^0_+,~~~\tilde{\pi}_+ = e^{-\zeta \gamma^{34}}\tilde{\pi}^0_+,
\eeq
where we have introduced linearly independent constant spinors obeying the projections
\begin{align}\label{eq:pi0projections}
&\gamma^{1456}\pi^0_+= -\pi^0_+,~~~\gamma^{2345}\pi^0_+=\pi^0_+,\nn\\[2mm]
&\gamma^{1456}\tilde\pi^0_+= -\tilde\pi^0_+,~~~\gamma^{2345}\tilde\pi^0_+=\tilde\pi^0_+,
\end{align}
in the frame of eq (\ref{eq:frame}). 
The 10d spinor is constructed as in eq (\ref{eq: MWSpinors}), but all dependence on the $\mathbb{CP}^3$ directions is in eq (\ref{eq: dependence}), which is clearly independent of the $SU(2)$ directions. 

So  $\mathcal{N}=2$ is preserved under the NAT duality transformation. We show in Appendices \ref{sec:SUSYdetailsABJM2},\ref{sec:SUSYdetailsABJMNATD} that the solution does this by mapping a $U(1)$'s worth of the $SU(3)$-structures supported by $\mathbb{CP}^3$ to a $U(1)$'s worth of dynamical $SU(2)$-structures defined on the dual internal space $\hat{M}_6$.  Of course only two of these dual objects are truly distinct: those that are defined in terms of the two linearly independent Killing spinors on $\hat{M}_6$. These may be summed to give the internal part of the $\mathcal{N}=2$ Killing spinor in IIB, namely
\begin{align}\label{eq: NATDspinor}
\hat{\eta}^1_+ &= e^{i\frac{3\pi}{4}}\bigg(e^{\zeta \gamma^{34}}\pi^0_++  e^{-\zeta \gamma^{34}}\tilde{\pi}^0_+\bigg),\nn\\[2mm]
\hat{\eta}^2_+&= e^{i\frac{3\pi}{4}}\bigg(\kappa_{\|}\big(e^{\zeta \gamma^{34}}\pi^0_++e^{-\zeta \gamma^{34}}\tilde{\pi}^0_+\big)+\kappa_{\perp}\mathcal{J}\big(e^{-\zeta \gamma^{34}}\pi^0_-+ +e^{\zeta \gamma^{34}}\tilde{\pi}^0_-\big)\bigg),
\end{align}
where $\kappa_{\perp}$ and $\kappa_{\|}$ satisfy $\kappa_{\perp}^2+\kappa_{\|}^2=1$  and are given in eq (\ref{eq: kappas}). The Matrix $\mathcal{J}$ is defined as
\beq
\mathcal{J}=i\frac{e^{C_1+C_2+C_3}\cos2\zeta \gamma^1+e^{C_1}v_1\gamma^4+e^{C_2}v_2 \gamma^5+e^{C_3}v_3\gamma^6}{\sqrt{ e^{2(C_1+C_2+C_3)}\cos^22\zeta+e^{2C_1}v_1^2+e^{2C_2}v_2^2+e^{2C_3}v_3^3}},
\eeq
in the frame of eq (\ref{eq:frame}), where $e^{C_a}$ are given in eq (\ref{NATDinput}). However for what follows it is only important that  $\mathcal{J}$ is independent of $\phi$.\\

It turns out that the amount of preserved supersymmetry is left invariant when we perform an additional T-duality on $\partial_{\phi}$. As proved in \cite{Hassan:1999bv} (see also \cite{Kelekci:2014ima}), to see this it is sufficient to show that the Killing spinors are independent of $\phi$ in the canonical frame of T-duality, where $\phi$ only appears in the vielbein
\beq
e^{\phi}= e^{C(x)}\big(d\phi+A(x)).
\eeq
We get to such a frame by performing a $SO(4)$ transformation $\mathcal{R}$ in eq (\ref{eq:framerotation}),  on the canonical NAT dual vielbeins in eq (\ref{eq:frame2}). The action of this rotation on the 10d Majorana Killing spinor will be $\epsilon \to \mathcal{S}\epsilon$, where
\beq
\mathcal{S}^{-1}\gamma^a \mathcal{S}= \mathcal{R}^a_{~b}\gamma^b.
\eeq
The matrix $\mathcal{S}$ will be complicated but will not depend on $\phi$ because $\mathcal{R}$ does not, which is all that matters. This together with the fact that eq (\ref{eq: NATDspinor}) is $\phi$ independent ensures all supersymmetry must be preserved. Indeed in the next section (see also Appendix \ref{sec:Mtheoryspinors}) we see that upon lifting to M-theory the NAT-T dual solution preserves $\mathcal{N}=2$ supersymmetry in the form of a local $SU(2)$ structure in 7d of the form given in \cite{Gabella:2012rc}.

\subsection{Properties of the CFT}

In this section we briefly discuss some properties of the CFT associated to the NAT-T dual
solution. The discussion follows very closely the analysis in reference \cite{Lozano:2014ata} for the IIB NAT solution. For this reason we will omit most of the explicit computations. The reader is referred to this reference for more details.

It was shown in  \cite{Lozano:2013oma, Lozano:2014ata} that the presence of large gauge transformations in NAT dual backgrounds allows to constrain quite non-trivially their global properties. In our particular background (see \cite{Lozano:2014ata})
it is easy to see that at  the singularity $\zeta=0$ the NS 2-form given by (\ref{eq:B2_NATD_T}) reduces to
\beq
B_2 \lvert_{\zeta\sim 0}= - r {\rm Vol}(S^2)\, ,
\eeq
while the space spanned by $(\zeta,S^2)$ becomes conformal to a singular cone with boundary $S^2$. Therefore large gauge transformations can be defined on this non-trivial 2-cycle, which must render 
\begin{equation}
b=\frac{1}{4\pi^2}\left| \int_{S^2} B_2 \right|
\end{equation}
in the fundamental region. For this, $B_2$ must transform into $B_2\rightarrow B_2+n\pi {\rm Vol}(S^2)$ when $r\in [n\pi, (n+1)\pi]$.

In turn, the ${\hat F}_3$ and ${\hat F}_5$ field strengths lying on the $\zeta, \theta, \phi$ and $\zeta, \theta, \phi, S^2$ directions of the NAT dual solution in \cite{Lozano:2014ata} give rise after the T-duality to ${\hat F}_2$ and ${\hat F}_4$ field strengths lying on the $\zeta, \theta$ and $\zeta, \theta, S^2$ directions, with the second one non-vanishing only in the presence of large gauge transformations. Accordingly, a Page charge associated to the $\zeta, \theta$ components of ${\hat F}_2$ is generated in IIA from the quantization condition
\begin{equation}
\frac{1}{2\kappa_{10}^2}\int {\hat F}_p=T_{8-p}N_{8-p}\, .
\end{equation}
As in \cite{Lozano:2014ata} this charge is to be interpreted as the rank of the gauge group of the CFT dual to the solution in the $r\in [0,\pi]$ interval. We indeed get $N_6=N_5$, with $N_5$ given by (3.18) in  \cite{Lozano:2014ata}. Specifically this fixes $L$ to satisfy
\beq\label{eq: defN}
k L^4= 64\pi N_6.
\eeq
Note that this means that color branes are now D6-branes 
spanned on the $\mathbb{R}_{1,2}\times M_1\times S^1_\phi \times S^2$ directions. One can indeed check that these branes are BPS when placed at the $\zeta=0$ singularity. As in \cite{Lozano:2014ata} the combination $e^{-\phi}\sqrt{g_{rr}}$ in the DBI action is non-singular, rendering well-defined color branes. 

In the presence of large gauge transformations with parameter $n$ there is also a non-vanishing D4-brane charge $N_4=n N_6$, equal to the $N_3$ charge in \cite{Lozano:2014ata}. Indeed one can check that D4-branes are also BPS when placed at $\zeta=0$. These branes should also play a role as color branes for $n\neq 0$, that is, in the $r\in [n\pi, (n+1)\pi]$ intervals. The physical interpretation is that 
$N_4$ charge is created in the worldvolume of a D6 when it crosses $n$ NS5. It is then plausible that the field theory dual to the solution in the $[n\pi, (n+1)\pi]$ intervals
arises in a (D4, D6) bound state - NS5 intersection. A similar realization was suggested in \cite{Bea:2015fja} for $AdS_3$ duals.

The charge interpreted as level in ABJM is also doubled under the NAT duality transformation. As a result, after the new Abelian T-duality we get two charges, $k_6$, $k_4$, associated to the $(r,\theta)$ and $(r,\theta, S^2)$ components of ${\hat F}_2$ and ${\hat F}_4$, respectively. They thus correspond to D6 and D4 branes or, equivalently, to D4-branes carrying both monopole and dipole charges. We may express the levels in terms of $k$ and the number of large gauge transformations performed as
\beq\label{eq: defk}
k_6= k \frac{(2n+1)\pi}{4},~~~ k_4 = k \frac{(3n+2)\pi}{12}\, .
\eeq

Finally, it is easy to check that particle-like brane configurations can be associated to each of the charges with an interpretation as either rank or level in the IIA background. These branes are in all cases D2 or D4 branes wrapped on different sub-manifolds of the internal space. In particular:
\begin{itemize}
\item Di-monopoles $\leftrightarrow$ D2 on $M_1 \times S^1_\phi$, D4 on $M_1 \times S^1_\phi \times S^2$
\item 't Hooft monopoles $\leftrightarrow$ D2 on $\{M_1, \theta\}$, D4 on $\{M_1, \theta, S^2\}$
\item Di-baryons $\leftrightarrow$ D2 on $\{\zeta, S^1_\phi\}$, D4 on $\{\zeta, S^1_\phi, S^2\}$
\item Baryon vertices $\leftrightarrow$  D2 on $\{\zeta, \theta\}$, D4 on  $\{\zeta, \theta, S^2\}$
\end{itemize}
As for the IIB $AdS_4$ solution (see \cite{Lozano:2014ata} for the details) the di-monopoles and 't Hooft monopoles have to sit at $\zeta=0$ while the di-baryons sit at $r=0$.

The previous analysis suggests a dual CFT in the $r\in [0,\pi]$ region defined in terms of a $U(N_6)_{k_4}\times U(N_6)_{-k_4}$ quiver gauge theory with ${\mathcal N}=2$ supersymmetry, sourced by D6-branes spanned on the $\mathbb{R}_{1,2}\times M_1\times S^1_\phi\times S^2$ directions\footnote{One can see (see  \cite{Lozano:2014ata}) that $k_4$ is the level associated to the D6 color branes.}. In turn, for $r\in [n\pi, (n+1)\pi]$ the gauge theory would arise from 
(D4, D6) - NS5 intersections. It was argued in \cite{Lozano:2014ata}
 that invariance under large gauge transformations would imply that the seemingly different CFTs dual to the solution as the non-compact internal direction increases, could be related by some kind of duality, as in \cite{Benini:2007gx}, with the essential difference that in this case the flow parameter would not be the energy scale but the non-compact internal direction. Reference \cite{Bea:2015fja}
proposed an alternative mechanism which, applied to our solution, would imply that new $U(N_6)\times U(N_6)$ gauge groups would be created by some kind of un-higgsing mechanism, also not related to an energy scale, every time a NS5-brane is crossed. It would be interesting to understand better these proposals for the dual CFT as $r$ increases. 
 
In any case, keeping in mind that there is no a priori reason to expect that the new geometry makes sense as a string theory background\footnote{An essential difference with respect to its Abelian counterpart is that non-Abelian T-duality has not been proved to be a symmetry of String Theory (see \cite{Alvarez:1993qi}).}, we can just take these proposals as stringy inspired arguments in favor of the existence of a fundamental region in which the dual CFT would contain the same number of gauge groups as the original one. 

Restricting ourselves to the $r\in [0,\pi]$ region, a candidate brane realization of the dual CFT would then be the T-dual of the brane picture proposed in  \cite{Lozano:2014ata} for the NAT dual of ABJM:

\begin{equation}
 \begin{array}{cc|lclclclclclclclc}
 \label{system1}
5_2^2:\ \ \ \    \  &    \times & \times &    \times &  \times &   \times  & \times &  -  & z_1 &  z_2 &   - \\
N_6\, D6:\ \ \ \   \  &  \times & \times &  \times & -  & \times & -  & \times  & \times &  \times  & -   \\
(5_2^2,k_4\,  {\rm D4}): \ \  &    \times & \times & \times & - & \times &  \cos{\theta}  & - & - & - & \sin{\theta}  
                         \end{array} 
\end{equation}
where $5^2_2$ denotes the IIA exotic brane that arises after a T-duality transformation along a worldvolume direction of the $5^2_2$ exotic brane of the IIB configuration \cite{Lozano:2014ata}, and 
$z_1$ and $z_2$ denote the two special Killing directions of this brane \cite{Hull:1997kb,Obers:1998fb}.
In our notation the 
$(5_2^2,k_4\,  {\rm D4})$ bound state would be extended along the 0124 and $x^5 \cos{\theta}+x^9 \sin{\theta}$ directions, and its relative orientation w.r.t. the $5^2_2$-brane in the 59 plane would depend on $k_4$.

Note that the previous picture implies that in M-theory the corresponding $AdS_4$ geometry would be sourced in the fundamental region $r\in [0,\pi]$ by Kaluza-Klein monopoles, as we discuss in the next section.

\section{The purely magnetic $AdS_4$ solution in M-theory}

\label{sec:lift_Mtheory}

In this section we lift the solution of the previous section to M-theory and show that it falls into the general class of solutions with purely magnetic flux considered in  \cite{Gauntlett:2006ux}. The analysis of quantized charges suggests a dual CFT arising from Kaluza-Klein monopoles and M5-branes wrapped on the Taub-NUT direction of the monopoles. We compute the central charge and show that it scales with $(N_5 + N_6/2)^{3/2}$, where $N_5$ is the number of wrapped M5-branes and $N_6$ the number of Kaluza-Klein monopoles. This becomes simply $N_6^{3/2}$ in the fundamental region $r\in [0,\pi]$. 

\subsection{Fluxes}

The RR potentials of the IIA solution are given by
\beq\label{eq:C1}
C_1 =\frac{k}{16}\bigg(\cos^2\zeta\cos\theta(3L^4\sin^3\zeta\cos\zeta d\zeta -8 r dr)-8r \cos\chi \sin^2\zeta d\phi\bigg),
\eeq
\beq
\label{C3hat}
C_3 - B_2 \wedge C_1 = \frac{k}{2} r^2 \cos\zeta \sin\chi \bigg(\sin\zeta \sin\chi d\zeta\wedge d\xi\wedge d\phi-\cos\theta\cos\zeta d\xi\wedge dr\wedge d\chi\bigg).
\eeq

$C_1$ gives rise to the $g_{\mu z}/g_{zz}$ component of the 11d metric, where $z$ denotes the eleventh direction. Given that there is a magnetic charge associated to $C_1$ in IIA,
a quantized Taub-NUT charge arises in 11d. The brane that carries Taub-NUT charge is the Kaluza-Klein monopole, which is connected to the IIA D6-brane upon reduction along the eleventh, Taub-NUT direction. Since the IIA solution was sourced by D6-branes in the fundamental region $r\in [0,\pi]$, Kaluza-Klein monopoles should play the role of color branes in M-theory in this region.
As we will discuss, BPS KK-monopoles spanned on the $\mathbb{R}_{1,2}\times M_1\times S^1_\phi\times S^2$ directions  can indeed be constructed in 11d that give rise to the D6 color branes in IIA upon reduction.

$(C_3-B_2\wedge C_1)$ gives rise in turn to the 3-form\footnote{Our notation is that $i_k C_3$ denotes the interior product of $C_3$ with the Killing vector $k^\mu=\delta^\mu_z$, that points on the eleventh direction, $k_1$ is the 1-form $k_1=i_k g$ and $k^2$ the scalar $k^2=i_k i_k g$, where $g$ stands for the eleven dimensional metric.}

\begin{equation}
\label{C_3trans}
\hat{C}_3=C_3-i_k C_3 \wedge \frac{k_1}{k^2}
\end{equation}
in 11d. Note that $\hat{C}_3$ has no components along the eleventh direction. This will be of relevance in our later discussion. 
The M-theory 4-form flux is then given by
\begin{equation}
G_4=dC_3=d\Bigl( {\hat C}_3+i_k C_3 \wedge (\frac{k_1}{k^2}+dz)\Bigr)
\end{equation}
which, as we can see, is purely magnetic. Therefore there will be no M2-branes sourcing the 11d solution. 

As we have noted, $\hat{C}_3$ is by construction transverse to the eleventh direction. This potential couples in the worldvolume of M2-branes constrained to move in the space transverse to the Killing direction and in the worldvolume of M5-branes wrapped on this direction \cite{Lozano:2000aq}.
Moreover, its magnetic components are associated to wrapped M5-branes. Indeed one can show that these branes are  BPS in the 11d background and are to be interpreted as color branes. They give rise upon reduction  to the color D4-branes of the IIA background. Other field theory observables that we will be able to describe holographically will be constructed in terms of M2-branes transverse to the Killing direction or M5-branes wrapped on this direction.

\subsection{Geometry and local $SU(2)$ structure}

\label{sec: SU2}
In \cite{Gabella:2012rc} it was shown that the most general $\mathcal{N}=2$ preserving solution in M-theory with an $AdS_4$ factor supports an $SU(2)$ structure in 7d. As the  M-theory 4-form $G_4$ is purely magnetic it actually falls into the more constrained class of solutions originally considered in \cite{Gauntlett:2006ux}. In this section we show that we can uplift the IIA solution to M-theory and fit it into this class of solutions.

The metric ansatz of  \cite{Gabella:2012rc} is of the form
\beq
ds^2_{11}=e^{2\tilde{\Delta}}\bigg( ds^2 (AdS_4)+ ds^2(\mathcal{M}_7)\bigg)
\eeq 
where we have
\beq
e^{2\tilde{\Delta}}= L^2 e^{-\frac{2}{3}\Phi},~~~ ds^2(\mathcal{M}_7) = \frac{1}{L^2}\bigg[ ds^2 (\mathcal{M}_6)+ e^{2\Phi}\big(C_1+ dz\big)^2\bigg],
\eeq
so that $\text{Ricci}(AdS_4)=-12\, g(AdS_4)$ to match the conventions of \cite{Gabella:2012rc}. The metric on $\mathcal{M}_6$ is mearly the internal part of the IIA metric in eq (\ref{IIAmetric}).\\

It is possible to express the internal 7d metric in the form
\beq
ds^2(\mathcal{M}_7)= ds^2(SU(2))+ E_1^2+E_2^2+E_3^2
\eeq
where $ds^2(SU(2))$ is the metric on a 4 manifold supporting a canonical $SU(2)$-structure with associated real 2-form $J=J_3$ and holomorphic 2-form $\Omega=J_1+i J_2$, satisfying 
\beq
J_3\wedge J_3= \frac{1}{2}\Omega\wedge \bar{\Omega},~~~J_3\wedge\Omega=0,~~~ \iota_{E_i}J_3=\iota_{E_i}\Omega=0.
\eeq
Since  $G_4$ is purely magnetic it is possible to define local coordinates such that \cite{Gauntlett:2006ux}
\beq
E_1 = \frac{1}{4}e^{-3\tilde{\Delta}} \rho \,d\xi,~~ E_2 = \frac{1}{4} e^{-3\tilde{\Delta}} \frac{d\rho}{\sqrt{1- e^{-6\tilde{\Delta}} \rho^2}},
\eeq
where $\xi$ parametrizes the $U(1)$ R-symmetry and $\rho$ is defined through the associated Reeb vector $\tilde{\xi}$ as $|\tilde{\xi}|^2=e^{-6\Delta} \rho^2$. Supersymmetry then requires that the $SU(2)$ forms satisfy
\begin{align}
d\big( e^{3\tilde{\Delta}} \sqrt{1-|\tilde{\xi}|^2}E_3\big)&=e^{3\tilde{\Delta}}\big(2 J_3-2 |\tilde{\xi}| E_2\wedge E_3\big)\nn,\\[2mm]
d\big(|\tilde{\xi}|^2 e^{9\tilde{\Delta}} J_2\wedge E_2\big)&= e^{3\tilde{\Delta}}|\tilde{\xi}| d\big(e^{6\tilde{\Delta}} J_1\wedge E_3\big),\nn\\[2mm]
d\big(e^{6\tilde{\Delta}}J_1 \wedge E_2\big)&=- e^{3\tilde{\Delta}}|\tilde{\xi}| d\big(e^{3\tilde{\Delta}} J_2\wedge E_3\big),
\end{align}
and the flux be given by
\beq
G_4 = \frac{1}{4}d\xi\wedge d\big( e^{3\tilde{\Delta}}\sqrt{1- |\tilde{\xi}|^2} J_1\big).\\
\eeq

We find that the uplift of the IIA solution fits into the above parametrisation. All the forms are defined in terms of the internal M-theory Killing spinors derived in  Appendix \ref{sec:Mtheoryspinors}, one needs only to plug them into the bi-linears in Appendix B of \cite{Gabella:2012rc}. Performing these steps with some liberal application of Mathematica, we find the local coordinate 
\beq
\rho = \frac{k}{8} L^4 y_1\sin^22\zeta \sin\theta,
\eeq
and the solutions specific vielbein
\begin{align}
E_3 = -\frac{e^{-3\tilde{\Delta}}}{\sqrt{1- e^{-6\tilde{\Delta}}\rho^2}}\bigg[&\frac{kL^2}{256}\sin^22\zeta\bigg(L^4\sin^22\zeta\sin\theta d\theta+64 y_2d\phi-64(y_1 dy_1+y_2 dy_2)\cos\theta\bigg)+\nn\\
&L^2 \cos2\zeta \big(dz+C_1\big)\bigg],
\end{align}
where $C_1$ is the potential giving rise to the IIA RR 2-form, which may be found in eq (\ref{eq:C1}).
To express the $SU(2)$ forms we introduce the following orthonormal frame
\begin{align}
e^1=&\frac{1}{\sqrt{X_1}\sqrt{X_2}}\bigg(2 X_1 dy_2+ 32y_1 y_2 \cos^2\zeta\sin\theta\big(\sin\theta dy_1+y_1 \cos\theta d\theta\big)\bigg),\\[3mm]
e^2= &\frac{e^{-3\tilde{\Delta}}k L^2}{8 \sqrt{X_1}\sqrt{X_2}}\bigg(\cos^2\zeta \cos\theta\big(X_1(16 y_2^2+L^4 \sin^4\zeta)dy_2 -256 y_1^3 y_2\sin^2\theta dy_1\big) -X_2\sin^2\zeta d\phi\bigg)-\nn\\[2mm]
&\frac{32 y_2e^{3\tilde{\Delta}}}{kL^6 \sin^2\zeta \sqrt{X_1}\sqrt{X_2}}\bigg(16y_2 \cos\theta dy_2-L^4\cos^2\zeta\sin^2\zeta \sin\theta d\theta\bigg),\nn\\[3mm]
e^3= &\frac{e^{-3\tilde{\Delta}}k L^4\sin2\zeta}{32\sqrt{X_1}\sqrt{1-e^{-6\Delta}\rho^2}}\bigg(-X_1 d\zeta+4 y_1\sin2\zeta \cos2\zeta\sin\theta \big(\sin\theta dy_1+y_1 \cos\theta d\theta\big)\bigg),\nn\\[3mm]
e^4=&\frac{\cos 2\zeta}{2L^2\sqrt{X_1}}\bigg(16y_2 d\phi+L^4 \cos^2\zeta\sin^2\zeta\sin\theta d\theta-16\cos\theta\big(y_1dy_1+y_2dy_2\big)\bigg)-\nn\\
&\frac{k L^6 \cos^2\zeta \sin^2\zeta \sqrt{X_1}e^{-3\tilde{\Delta}}}{8 \sqrt{1-e^{-6\Delta} \rho^2}}\bigg(dz+C_1\bigg)\nn,
\end{align}
where
\begin{align}
X_1&=16y_1^2\cos^2\theta \sin^2\zeta+16 y_2^2sin^2\theta\cos^2\zeta+L^4\sin^2\theta\sin^4\zeta \cos^2\zeta ,\\
X_2&=16y_1^2\cos^2\theta\big(16y_1^2+L^4 \sin^4\zeta\cos^2\zeta\big)+L^4\cos^2\zeta\sin^2\zeta\sin^2\theta\big(16 y_1^2+\cos^2\zeta(16y_2^2+L^4 \sin^4\zeta)\big)\nn.
\end{align}
With respect to this basis we have
\beq
J= e^1\wedge e^2+e^3\wedge e^4,~~~\Omega= e^{i\alpha}(e^1+i e^2)\wedge (e^3+i e^4),
\eeq
where the phase $\alpha$ is defined through
\beq
\tan\alpha= -\frac{e^{3\tilde{\Delta}}}{k L^2 \cos\theta \,y_1^2}.
\eeq

\subsection{Properties of the CFT}

As in the previous section, some properties of the CFT dual can be inferred by analyzing the 11d supergravity solution. The picture that arises is simply the 11d realization of the IIA
picture described in subsection 3.2, apart from some subtleties that have to do with the existence of the special Taub-NUT direction. Indeed, all brane configurations that play a role in 11d will be either transverse to this direction or wrapped on it.

The non-trivial $S^2$ of the IIA geometry is also present in the 11d uplift. Therefore one can define large gauge transformations for the uplift of the $B_2$ field, which is the 11d 3-form potential with a component along the Taub-NUT direction, $i_k C_3$. Thus, as in the IIA background, we need to divide $r$ in intervals of length $\pi$ in order to have $i_k C_3$ lying in the fundamental region. From here the discussion parallels exactly the IIA discussion.

In 11d we find quantized charges $N_6$ and $N_5=n N_6$, equal to the $N_6$ and $N_4$, respectively, in IIA. $N_6$ is associated to KK-monopoles and $N_5$ to M5-branes wrapped on the Taub-NUT direction of the monopole. The interpretation is that M5-brane charge (with the M5-brane wrapped in the Taub-NUT direction of the monopole) is created in the worldvolume of the KK-monopole when it crosses M5-branes transverse to the Taub-NUT direction\footnote{Recall that in 11d $i_k C_3\rightarrow i_k C_3+n\pi {\rm Vol} (S^2)$, and the M5 is magnetically charged with respect to this field.}. Using the worldvolume effective action that describes a KK-monopole in 11d \cite{Bergshoeff:1997gy,Bergshoeff:1998ef} one can easily check that it is BPS when placed at $\zeta=0$. The calculation parallels the D6-brane calculation in IIA with the only difference that the action is now written in terms of eleven dimensional fields. Similarly an M5-brane wrapped on the Taub-NUT direction is also BPS at this location.

As in IIA, the charge interpreted as level in 11d is also doubled, and we get two values 
$k_6$ and $k_5$ equal to the $k_6$ and $k_4$ charges, respectively, in IIA. These are now associated to wrapped M5-branes carrying monopole and dipole charges\footnote{As shown in \cite{Lozano:2000aq}, M5-branes wrapped on an isometric direction can carry KK-monopole charge, with the Taub-NUT direction equal to the isometric direction.}.

Similarly, we find particle-like brane configurations, which are either M2-branes transverse to the Taub-NUT direction or M5-branes wrapped on this direction. These branes are wrapped on the same sub-manifolds of the internal space as in IIA. Namely, 
\begin{itemize}
\item Di-monopoles $\leftrightarrow$ M2 on $M_1 \times S^1_\phi$, M5 on $M_1 \times S^1_\phi \times S^2\times S^1_z$
\item 't Hooft monopoles $\leftrightarrow$ M2 on $\{M_1, \theta\}$, M5 on $\{M_1, \theta, S^2, S^1_z\}$
\item Di-baryons $\leftrightarrow$ M2 on $\{\zeta, S^1_\phi\}$, M5 on $\{\zeta, S^1_\phi, S^2, S^1_z\}$
\item Baryon vertices $\leftrightarrow$  M2 on $\{\zeta, \theta\}$, M5 on  $\{\zeta, \theta, S^2, S^1_z\}$
\end{itemize}
As for the IIB $AdS_4$ solution (see \cite{Lozano:2014ata} for the details) the di-monopoles and 't Hooft monopoles have to sit at $\zeta=0$ while the di-baryons sit at $r=0$. In these derivations we have used the action that describes M2-branes transverse to the Taub-NUT direction of the monopole. In this action $i_k C_3$ couples in both the DBI and CS parts, so the M2-branes are sensitive to large gauge transformations. The details of this action can be found in \cite{Lozano:2000aq}.

Putting together this information, and in analogy with the IIA discussion, we expect a field theory in the $r\in [0,\pi]$ region described by a $U(N_6)_{k_5}\times U(N_6)_{-k_5}$ quiver with $\mathcal{N}=2$ supersymmetry, sourced by KK-monopoles spanned on the $\mathbb{R}_{1,2}\times M_1\times S^2\times S^1_\phi$ directions, and with Taub-NUT direction $z$. A possible brane realization in the fundamental region $r\in [0,\pi]$ could be
\begin{equation}
 \begin{array}{cc|lclclclclclclclc|c}
 \label{system2}
5^3:\ \ \ \    \  &    \times & \times &    \times &  \times &   \times  & \times &  -  & z_1 &  z_2 &   - & z \\
N_6\, M6:\ \ \ \   \  &  \times & \times &  \times & -  & \times & -  & \times  & \times &  \times  & - & z  \\
(5^3,k_4\,  {\rm M5}): \ \  &    \times & \times & \times & - & \times &  \cos{\theta}  & - & - & - & \sin{\theta} & \times 
                         \end{array} 
\end{equation}
where $z$ denotes the eleventh direction, the M6 is the Kaluza-Klein monopole with Taub-NUT direction $z$ and $5^3$ is the exotic brane that gives rise to the IIA $5^2_2$ brane upon reduction \cite{Hull:1997kb,Obers:1998fb}. 

\subsection{Free energy}

We can now calculate the free energy on a 3-sphere in the CFT dual to the solution in M-theory. This is expressed in terms of the effective 4 dimensional Newton constant $G_4$ as
\beq
\mathcal{F}_{S^3} = \frac{\pi}{2 G_4}.
\eeq
One can determine $G_4$ via a dimensional reduction of supergravity on the internal space $\mathcal{M}_7$, the result is
\beq
\frac{1}{16\pi G_4} = \frac{\pi}{2 (2\pi)^9}\int_{\mathcal{M}_7} e^{9\tilde{\Delta}} {\rm Vol} (\mathcal{M}_7),
\eeq 
where we work in units such that $ l_p=1$. For the case at hand the relevant quantity is
\beq
e^{9\tilde{\Delta}}{\rm Vol}(\mathcal{M}_7) = \frac{k^2 L^6}{32}r^2 \sin^3\zeta\cos^3\zeta \sin\theta\sin\chi d\zeta\wedge d\theta\wedge d\phi \wedge dr\wedge d\chi\wedge d\xi\wedge dz.
\eeq
Integrating this in the region $r\in [n\pi, (n+1)\pi]$, $z\in [0,2\pi]$ and using eqs (\ref{eq: defN},\ref{eq: defk}) we arrive at
\beq
\mathcal{F}_{S^3} =  \frac{\sqrt{2}\pi}{36}\left(12+ \frac{N_6^2}{\left(N_5+\frac{N_6}{2}\right)^2}\right)\sqrt{\kappa_6}\left(N_5 + \frac{N_6}{2}\right)^{3/2}.
\eeq
This reproduces, as expected, the result in IIB, with $N_5,N_3\to N_6,N_5$ \cite{Lozano:2014ata}. Essentially we have $\mathcal{F}_{S^3}\sim\left(N_5 + \frac{N_6}{2}\right)^{3/2}$ which reproduces the $N^{3/2}$ behavior of ABJM. In particular, in the fundamental region $r\in [0,\pi]$, where $N_5=0$, we find
\begin{equation}
\label{freee}
\mathcal{F}_{S^3}=\frac{\sqrt{2}\pi}{3^{3/2}}\sqrt{k_4}N_6^{3/2}\, .
\end{equation} 

This is not a surprising result, given that the dependence of the free energy in type II theories, like the central charge and entanglement entropy of the strip, depends on the internal directions only through the quantity
\beq
V_{int} =\int_{\mathcal{M}_6} e^{-2\Phi} {\rm Vol}(\mathcal{M}_6),
\eeq
and is thus invariant under Abelian T-duality\footnote{It is not invariant though under non-Abelian T-duality, because even if the integrand is invariant, the $S^3$ on which the dualisation is performed is transformed into an $M_1\times S^2$ space, where $M_1$ is the space spanned by the $r$-direction, and thus the domain of integration changes. This is the reason why the prefactors in (\ref{freee}) are not the same as in ABJM.} and uplift to 11d.

So, quite surprisingly, we have found an $AdS_4$ M-theory solution with purely magnetic flux that falls in the general classification of \cite{Gauntlett:2006ux}, that originates in M5-branes wrapped on calibrated 3-cycles, but whose free energy does not exhibit the expected $N^3$ behavior. We leave a further discussion on this issue for the conclusions.

\section{Conclusions}
In this work we have presented a new warped $AdS_4$ solution of M-theory preserving $\mathcal{N}=2$ supersymmetry, giving the only known solution in this class other than the uplift of Pernici-Sezgin. A legitimate question to ask is whether this solution is truly distinct from Pernici-Sezgin, indeed this solution was generated by performing first a NAT then T duality on $AdS_4\times \mathbb{CP}^3$, and some geometries derived via NAT duality have been shown to fall within the ansatz of previous solutions. This does not seem to be the case with this example: The quickest thing to note is that the free energy of Pernici-Sezgin scales as $N^3$ while this solution scales as $N^{3/2}$. Additionally the uplift of Pernici-Sezgin is everywhere non singular while the curvature invariants of this solution blow up in certain regions of parameter space. One might still wonder if this solution approximates Pernici-Sezgin at least locally away from the singularity, as was argued in  \cite{Sfetsos:2010uq} to be the case for the NAT dual of $AdS_5\times S^5$  and the Gaiotto-Maldacena geometries \cite{Gaiotto:2009gz}. This also does not seem to be the case. Sfetsos and Thompson were able to find an additional solution to the Gaiotto-Maldacena Toda equation which gave their solution. The differential equations giving rise to Pernici-Sezgin are more simple and are solved uniquely. So this solution is truly distinct.

In this work, following on \cite{Lozano:2014ata}, we have taken the view that the range of $r$ is restricted to lie in a specific cell of length $\pi$ after $n$ large gauge transformations of $B_2$. The reason is to ensure that $0<\lvert\int_{S^2}B_2\lvert< 4\pi^2$, a restriction motivated by string theory. However this does present an issue for the geometry, we are choosing to end it at a regular point which would usually demand the inclusion of extra localized sources. From a purely geometric view point we might choose to take $0<r<\infty$, however this would be very undesirable from an AdS/CFT perspective. A continuous $r$ would lead to, among other things, a CFT dual with operators of continuous conformal dimension \cite{Lozano:2013oma}. An attractive resolution to these issues is that the NAT duality generates a solution which approximates a better defined solution free of these pathologies. At any rate, regardless of these potential criticisms, it seems likely that one could use this work as a stepping stone to further populate the solution space of purely magnetic M-theory solutions.

Supersymmetric probe branes in the 11d uplift of the Pernici-Sezgin solution were considered in \cite{Bah:2014dsa}, with an aim at introducing punctures on the Riemann surface along the lines of \cite{Gaiotto:2009gz}. The BPS configurations were shown to preserve two $U(1)'s$, one more than required by the R-symmetry of the 3d $\mathcal{N}=2$ SCFT. This second $U(1)$ corresponds to a global $U(1)$ in the 3d field theory, and seems to play a key role in the 3d-3d correspondence \cite{Chung:2014qpa}. 
It was argued in \cite{Bah:2014dsa} that a large number of supersymmetric M5's would ultimately backreact on the Pernici-Sezgin geometry to produce a new $AdS_4$ solution with a $U(1)^2$ isometry. It would be interesting to show whether the $AdS_4$ solution obtained in this paper, containing a $U(1)^2$ isometry, could be related to this physical situation.

That the free energy of our purely magnetic $AdS_4$ solution scales like $\mathcal{F}_{S^3}\sim N^{3/2}$ rather than $N^3$ is a little puzzling. It was proved in \cite{Gabella:2012rc} that the presence of M2 branes, whether accompanied by $M5's$ or not, always gives rise to $N^{3/2}$ behavior. However we know that our solution cannot contain M2 branes, indeed it is not possible to accommodate M2 branes in a purely magnetic flux ansatz, so what are we to make of this apparent contradiction. Firstly it should be noted that, at least as far as the authors are aware, there is no proof $\mathcal{F}_{S^3}\sim N^{3}$ holds universally for all wrapped M5 brane solutions. However this seems like an inadequate evasion of a confusing result. More likely is that the solution we present does not correspond to wrapped M5 branes.  Indeed, the ansatz taken in \cite{Gauntlett:2006ux} to derive the purely magnetic solutions is defined by requiring the Killing spinors to satisfy the same projection conditions as the wrapped branes. Yet the solutions need not describe in general M5-branes wrapped in 3d manifolds in the near horizon limit.  The metric we have obtained is rather complicated and it seems difficult to identify a 3-cycle in the internal geometry that such branes might wrap. This together with the fact that the free energy does not scale with $N^3$ is suggesting that this is indeed the case for our solution.

On the other hand, even if the CFT interpretation of the solution is yet very preliminary, we seem to have found that there are quantized charges associated to both KK-monopoles and M5-branes, with the first being the only sources of the geometry in the $r$-region that we have defined as the fundamental region. This is also suggestive of a geometry not originating from wrapped M5-branes.

Finally, let us comment on something slightly tangential. In the process of discussing the supersymmetry preserved by purely magnetic M-theory solutions we analised the G-structure preserved by the NAT dual of ABJM. We showed in appendix \ref{sec:SUSYdetailsABJMNATD} that this IIB solution preserves a $U(1)$'s worth of dynamical $SU(2)$-structures in 6d. We note that, it is possible to take the intersection of two of these and define an identity structure. However, given that a complete systematic study of $AdS_4$ solutions to type II supergravity preserving $\mathcal{N}=2$ supersymmetry is currently absent form the literature, we have not pursued this here. Even so we know that, as with the better studied $AdS_5$, $\mathcal{N}=1$ cases \cite{Gauntlett:2005ww,Apruzzi:2015zna},   supersymmetry should be preserved in terms of either a local ``$SU(2)$-structure'' or ``identity structure'' on the internal co-dimensions of the isometry dual to the $U(1)$ R-symmetry. The NAT dual of ABJM will certainly fall into the latter class.

\subsection*{Acknowledgements}
We would like to thank Jerome Gauntlett, Dario Martelli, Achilleas Passias, Alessandro Tomasiello and Alberto Zaffaroni for useful discussions. This work has been partially supported by the COST Action MP1210 ``The String Theory Universe''. 
Y.L. and J.M. are partially supported by the Spanish Ministry of Economy and Competitiveness grant FPA2012-35043-C02-02. J.M. is supported by the grant BES-2013-064815 of the same institution.
He is grateful for the warm hospitality extended by the theoretical physics group at Milano-Bicocca U. where part of this work was done.
 N.T.M is supported by INFN and by the European Research Council under the European Union's Seventh Framework Program (FP/2007-2013) - ERC Grant Agreement n. 307286 (XD-STRING).

\appendix

\section{Some details on the NAT and T duality transformations}

In this Appendix we give some details on the derivation of the solution in section \ref{sec: newIIASolution}. The starting point is the $AdS_4 \times \mathbb{CP}^3$ metric written as a Hopf fibration
\beq
ds^2= ds^2(M_7)+ e^{2C_a}(\omega_a+A_a)^2,
\eeq
where $\omega_a$ are $SU(2)$ left-invariant 1-forms satisfying $d\omega_a=\frac{1}{2}\epsilon_{abc}\omega_b\wedge \omega_c$ and
\begin{align}\label{NATDinput}
 ds^2(M_7) &= \frac{L^2}{4}\bigg[ ds^2(AdS_4)+ 4 d\zeta^2+\cos^2\zeta\big(d\theta^2+\cos^2\theta d\phi^2\big)\bigg],\nn\\[2mm]
e^{2C_1}&=e^{2C_2} = \frac{L^2}{4}\sin^2\zeta,~~~e^{2C_3} = \frac{L^2}{4}\sin^2\zeta\cos^2\zeta,\nn\\[2mm]
A_1 &= A_2 = 0,~~~A_3= \cos\theta d\phi,
\end{align}
where the $AdS$ radius is 1. Specifically we introduce the vielbeins
\begin{align}\label{eq:frame}
&e^{x^{\mu}}=\frac{L}{2}\rho\, dx^{\mu},~~~ e^{\rho}=\frac{L}{2\rho}d\rho,~~~ e^1= L d\zeta,~~~ e^2= \frac{L}{2}\cos\zeta d\theta,~~~e^3 = \frac{L}{2}\cos\zeta\sin\theta d \phi,\nn\\
& e^4 = \frac{L}{2}\sin\zeta\,\omega_1,~~~e^5 = \frac{L}{2}\sin\zeta\,\omega_2,~~~e^6 = \frac{L}{2}\sin\zeta\cos\zeta(\omega_3+\cos\theta d\phi).
\end{align}
The dilaton of this solution is constant and set to $e^{\Phi}=\frac{k}{L}$, while the non trivial fluxes are
\begin{align}
F_2 =&G_2+ J^a_1\wedge\big(\omega_a+A_a\big)+\frac{1}{2}\epsilon_{abc}K^a_0\big(\omega_b+A_b\big)\wedge \big(\omega_c+A_c\big),\nn \\[2mm]
F_4 =& G_4+ K^a_3\wedge\big(\omega_a+A_a\big)+\frac{1}{2}\epsilon_{abc}M^a_2\big(\omega_b+A_b\big)\wedge \big(\omega_c+A_c\big)\nn\\
&+N_1\big(\omega_1+A_1\big)\wedge \big(\omega_2+A_2\big)\wedge \big(\omega_3+A_3\big),
\end{align}
where the only non zero components are
\begin{align}
& G_2 = -\frac{k}{2}\cos^2\zeta\sin\theta d\theta\wedge d\phi,~~~ J^3_1=-k \sin\zeta\cos\zeta d\zeta,~~~ K_0^3=-\frac{k}{2} \sin^2\zeta,\nn\\[2mm]
&G_4 = +\frac{3k L^2}{8} Vol(AdS_4).
\end{align}

\subsection{The IIB NAT Duality}

Expressing the solution in this manner allows one to simply read off the result of performing a NAT duality transformation on $\omega_a$ using \cite{Kelekci:2014ima}. The dual metric is given by
\beq
d\hat{s}^2 =  ds^2(M_7)+ \sum_{a=1}^3 \hat{e}^{a+3}
\eeq
We have introduced cylindrical polars for the dual coordinates
\beq
v_1= y_1 \cos\xi,~~~ v_2= y_1 \sin\xi,~~~ v_3 = y_2,
\eeq
and choose to express the dual canonical vielbeins $\hat{e}$ in a way that makes the residual $U(1)$ isometry given by $\partial_\xi$ explicit
\begin{align}\label{eq:frame2}
&\cos\xi\hat{e}^4+\sin\xi\hat{e}^5=-\frac{1}{8 L \Delta}\sin\zeta\bigg[4y_1 y_2\big(4 dy_2+L^2 \sin^2\zeta\cos^2\zeta(d\xi+\cos\theta d\phi)\big)\nn\\
&~~~~~~~~~~~~~~~~~~~~~~~~~~~~~~~~~~~~~~~~~~+ dy_1 \big(16y_1^2 +L^4 \sin^4\zeta \cos^2\zeta\big)\bigg],\nn\\[2mm]
&\cos\xi\hat{e}^5-\sin\xi\hat{e}^4=-\frac{1}{8 L  \Delta}\sin\zeta\bigg[4y_1dy_2\nn \\
&~~~~~~~~~~~~~~~~~~~~~~~~~~~~~~~~~~~~~~~~~~+\cos^2\zeta\big(-4y_2 dy_1+L^2 y_1\sin^2\zeta(d\xi+\cos\theta d\phi)\big)\bigg],\nn\\[2mm]
&\hat{e}^6=-\frac{1}{8 L \Delta}\sin\zeta\cos\zeta\bigg[16y_1y_2dy_1+dy_2(16y_2^2+L^4\sin^4\zeta)-4L^2y_1^2\sin^2\zeta(d\xi+\cos\theta d\phi)\bigg]\nn\\[2mm]
&\hat{e}^a =e^a,~~~ a= x^{\mu},\rho,1,2,3,
\end{align}
where we define
\beq
\Delta= \sin^2\zeta\big(y_1^2+\cos^2\zeta y_2^2 +\frac{L^4}{16}\sin^4\zeta\cos^2\zeta\big).
\eeq
A NS two form is generated
\beq
B_2=\frac{1}{\Delta}y_1\sin^2\zeta\bigg(y_1dy_2- y_2\cos^2\zeta dy_1\bigg)\wedge d\xi-\frac{1}{\Delta}\sin^2\zeta\cos^2\zeta \cos\theta\bigg(y_1y_2\,dy_1+\big(y_2^2+ \frac{L^4}{16}\sin^4\zeta\big)dy_2\bigg)\wedge d\phi  \nn
\eeq
while the dilaton becomes\footnote{Notice that, for simplicity in other expressions, we are extracting a factor of $\frac{L^2}{4}$ with respect to the definition of $\Delta$ in \cite{Kelekci:2014ima}.}
\beq
 e^{-2\hat{\Phi}}= \frac{L^2}{4}\,\Delta\, e^{-2\Phi}.
\eeq
The solution also has all possible RR forms turned on. These can be found in \cite{Lozano:2014ata} where this solution was originally derived.\\

\subsection{The IIA NAT-T Duality}

We would now like to perform a T-duality on the global $U(1)$ corresponding to $\partial_{\phi}$. To do this we can once more make use of the results of \cite{Kelekci:2014ima} (see \cite{Hassan:1999bv} for the original derivation). In order to do this we need to express the metric and $B_2$ as
\begin{align}
d\hat{s}^2&= d\hat{s}^2(M_9)+ e^{2C}(d\phi+A_1)^2,\nn\\[2mm]
B_2&= B+B_1\wedge d\phi.
\end{align}
Clearly $B_2$ is already in this form, while the same can be achieved for the metric with a rotation of the vielbein basis $\hat{e}\to \mathcal{R}\hat{e}$, giving
\begin{equation}
e^{2C}= \frac{\Xi}{4\Delta}L^2\cos^2\zeta,~~~A_1= \frac{y_1^2\sin^4\zeta \cos\theta}{\Xi}\,d\xi,
\end{equation}
where
\beq
\Xi= \Delta\sin^2\theta + y_1^2 \sin^4\zeta\cos^2\theta.
\eeq
A rotation that achieves this is 
\beq\label{eq:framerotation}
\mathcal{R}=\left(
\begin{array}{cccc}
 -\frac{\sqrt{\zeta _1^2+\zeta _2^2} \sin\zeta\cos \theta}{\sqrt{\Xi _0}} & \frac{\sin \theta \left(\sin \xi-\zeta _3
   \cos \xi\right)}{\sqrt{\Xi _0}} & -\frac{\sin \theta \left(\zeta _3 \sin \xi+\cos \xi\right)}{\sqrt{\Xi _0}} &
   \frac{\sqrt{\zeta _1^2+\zeta _2^2} \sin \theta}{\sqrt{\Xi _0}} \\
 0 & \frac{\zeta _3 \sin \xi+\left(\zeta _1^2+\zeta _2^2+1\right) \cos \xi}{\sqrt{\Delta _0} \sqrt{\zeta _1^2+\zeta _2^2+1}} &
   \frac{\left(\zeta _1^2+\zeta _2^2+1\right) \sin \xi-\zeta _3 \cos \xi}{\sqrt{\Delta _0} \sqrt{\zeta _1^2+\zeta _2^2+1}} &
   \frac{\sqrt{\zeta _1^2+\zeta _2^2} \zeta _3}{\sqrt{\Delta _0} \sqrt{\zeta _1^2+\zeta _2^2+1}} \\
 0 & -\frac{\sqrt{\zeta _1^2+\zeta _2^2} \sin \xi}{\sqrt{\zeta _1^2+\zeta _2^2+1}} & \frac{\sqrt{\zeta _1^2+\zeta _2^2} \cos (\xi
   )}{\sqrt{\zeta _1^2+\zeta _2^2+1}} & \frac{1}{\sqrt{\zeta _1^2+\zeta _2^2+1}} \\
 -\frac{\sqrt{\Delta _0} \sin \theta}{\sqrt{\Xi _0}} & \frac{\sqrt{\zeta _1^2+\zeta _2^2} \sin\zeta\cos \theta
   \left(\zeta _3 \cos \xi-\sin \xi\right)}{\sqrt{\Delta _0} \sqrt{\Xi _0}} & \frac{\sqrt{\zeta _1^2+\zeta _2^2} \sin\zeta
   \cos \theta \left(\zeta _3 \sin \xi+\cos \xi\right)}{\sqrt{\Delta _0} \sqrt{\Xi _0}} & -\frac{\left(\zeta _1^2+\zeta
   _2^2\right) \sin\zeta\cos \theta}{\sqrt{\Delta _0} \sqrt{\Xi _0}} \\
\end{array}
\right) 
\eeq
which acts on $2456$. We have introduced the following expressions
\beq
\Delta_0=1+\zeta_1^2+\zeta_2^2 + \zeta_3^2,~~~ \Xi_0 = \sin^2\theta \Delta_0+ \sin^2\zeta\cos^2\theta \big(\zeta_1^2+\zeta_2^2\big),~~~ \zeta_a= v_a e^{\sum_{b\neq a}C_a}.
\eeq
With the rotated vielbein basis, we may give the RR forms in \cite{Lozano:2014ata} in terms of them and then use  \cite{Hassan:1999bv,Kelekci:2014ima} to read off the T-dual solution, getting then the results in section \ref{sec: newIIASolution}.

\section{Type-II G-structure conditions for $AdS_4$ solutions}\label{sec: G-stucutureConvensions}
In this Appendix we review the G-structure conditions for supersymmetric $AdS_4\times M_6$  solutions, which is a slight modification\footnote{Specifically with the normalization of the internal spinor.} of what may be found in  \cite{Grana:2005sn,Grana:2006kf}, but with notation more akin to \cite{Andriot:2008va,Andriot:2010sya}. The metric  can be cast in the form
\beq
ds^2= e^{2A}ds^2( AdS_4)+ ds^2(M_6),
\eeq
where the $AdS$ radius is 1 and the dilaton has support only in $M_6$. The fluxes have the same direct product structure, which in terms of the RR polyform we may express as
\beq
F= F_{int}+ e^{4A} Vol(AdS_4)\wedge \tilde{F}.
\eeq
We use a real representation of the 10d gamma matrices\footnote{Specifically for the $AdS_4$ directions $\Gamma^{\mu}=  \hat{\gamma}^{\mu}\otimes 1$, while on $\mathbb{CP}^3$ we define $\Gamma^{i} = \gamma^{(4)}\otimes \gamma^i$, where $\hat{\gamma}^{\mu}$ and $\gamma^{a}$ are representations of the gamma matrices in 3+1 and 6 dimensions respectively. We define $\Gamma^{(10)}=\gamma^{(4)}\otimes \gamma^{(7)}$, where $\gamma^{(4)}= i\hat{\gamma}^{tx^1x^2 r}$ and $\gamma^{(7)} =-i \gamma_{123456}$.} in which the Dirac and ordinary conjugates coincide. 
A 4+6 split is performed on the 10d MW Killing spinors where $\epsilon= (\epsilon_1,\epsilon_2)^T$ and $\Gamma^{(10)}\epsilon=\sigma_3 \epsilon$ so that we can write
\begin{align}\label{eq: MWSpinors}
\epsilon_1 &= e^{A/2}\big(\zeta_+ \otimes \eta^{1}_++\zeta_- \otimes \eta^{1}_-\big),\nn\\
\epsilon_2 &= e^{A/2}\big(\zeta_+ \otimes \eta^{2}_{\mp}+\zeta_- \otimes \eta^{2}_{\pm}\big),
\end{align}
where $\pm$ labels chirality in 4 and 6 dimensions, so that the upper/lower signs should be taken in IIA/IIB and $(\eta_+)^* = \eta_-$ and we take the internal, $\eta^{1,2}$ spinor to have unit norm.\\

Preservation of supersymmetry may be expressed in terms of differential conditions on two pure spinors
\beq
\Psi_{\pm} = 8\, \eta^{1}_+\otimes \eta^{2\dag}_{\pm}.
\eeq
These conditions are given by
\begin{align}\label{eq: diffconSUSY}
(d-H)\wedge \big(e^{3A-\Phi}\Psi_{\pm}\big)&=-2 e^{2A-\Phi} \text{Re}\Psi_{\mp},\\[2mm]
(d-H)\wedge \big(e^{4A-\Phi}\Psi_{\mp}\big)&=-3e^{3A-\Phi}\text{Im}\Psi_{\pm} + e^{4A} \tilde{F}\nn,
\end{align}
where once more the upper/lower signs should be taken in IIA/IIB and 
\beq
e^{4A}\tilde{F} = \iota_{Vol(AdS_4)} F.\\
\eeq

The G-structure on $M_6$ can either be an $SU(3)$, when $\eta^{1}_+$ and $\eta^{2}_+$ are globally parallel, or $SU(2)$ when they are not. Using a Fierz identity and the Clifford map it is possible to express $\Phi_{\pm}$ as polyforms. In the $SU(3)$-structure case we can write this in terms of a complex 2-form $J$ and a holomorphic 3-form $\Omega_{hol}$ as 
\beq
\Psi_+ = e^{-i\theta_+} e^{-i J},~~~ \Psi_- = e^{i\theta_-}\Omega_{hol},
\eeq
where the forms may be expressed in terms of the internal spinors as 
\beq
J_{ab} = -i \eta^{\dag}_+ \gamma_{ab}\eta_+,~~~ \big(\Omega_{hol}\big)_{abc} = -\eta^{\dag}_-\gamma_{abc}\eta_+,
\eeq
where 
\beq
\eta^{1}_+ = e^{i \alpha_1 } \eta,~~~ \eta^{2}_+ = e^{i \alpha_2 } \eta,~~~ \eta^{\dag}_+\eta_+=1,~~~\theta_{\pm} =\alpha_1\mp \alpha_2,
\eeq
 and the forms obey $J\wedge J\wedge J = \frac{3i}{4}\Omega_{hol}\wedge \bar{\Omega}_{hol}$, $J\wedge \Omega_{hol}=0$.\\

For the $SU(2)$-structure case the internal spinor may be expressed as
\beq
\eta^1_+ = e^{i \alpha_1} \eta_+,~~~~ \eta^2_+ =e^{i \alpha_2} \big(\kappa_{\|} \eta_+ +\kappa_{\perp} \chi_+\big)
\eeq
where $\chi^{\dag}_+\eta_+=0$ and $\kappa_{\|}^2+\kappa_{\perp} ^2=1$. The pure spinors may then be expressed in terms of a holomorphic 1-form $z$, a real 2-form $j$ and a holomorphic 2-form $\omega_{hol}$ as
\begin{align}
\Psi_+&= i\, e^{i \theta_+} e^{\frac{1}{2}z\wedge \bar{z}}\wedge\bigg(\kappa_{\|} e^{-ij}- i \kappa_{\perp} \omega_{hol}\bigg)\nn,\\[2mm]
\Psi_-&=  e^{i \theta_-}z\wedge \bigg(\kappa_{\perp} e^{-i j}+ i \kappa_{\|} \omega_{hol}\bigg).
\end{align}
The various forms may be extracted from the spinor via
\beq
z_a = -i \eta^{\dag}_- \gamma_a \chi_+,~~~ j_{ab}= \frac{1}{2}\left( -i\eta^{\dag}_+\gamma_{ab} \eta_+ + i \chi^{\dag}_+\gamma_{ab}\chi_+\right),~~~ (\omega_{hol})_{ab}= +i \eta_-^{\dag}\gamma_{ab}\chi_-,
\eeq
and obey the conditions,
\beq
j\wedge j=\frac{1}{2}\omega_{hol}\wedge\bar{\omega}_{hol},~~~j\wedge\omega_{hol},~~~\omega_{hol}\wedge \omega_{hol}=0,~~~\iota_{z}\omega_{hol}=\iota_{z} j=0.
\eeq
Finally it should be noted that the above conditions are actually the conditions for $\mathcal{N}=1$ in 3d. We will be concerned with $\mathcal{N}=2$ supersymmetry which implies a CFT dual with $U(1)$ R-symmetry. This will manifest itself in the fact that there should be a $U(1)$'s worth of pure spinors satisfying eq (\ref{eq: diffconSUSY}), two of which are independent\footnote{In the sense that they can be constructed from two sets of linearly independent internal spinors $(\eta_1,\eta_2)$ and $(\tilde{\eta}_1,\tilde{\eta}_2)$.}.

\section{Detailed supersymmetry analysis}\label{sec:SUSYdetails}
In this appendix we shall look at how the Killing spinors are transformed under the sequence of dualities we perform to reach the M-theory solution of section \ref{sec: SU2}. We shall begin by identifying a set of spinors on $\mathbb{CP}^3$ that are uncharged under the $SU(2)$ on which the NAT- duality is performed.

\subsection{A $SU(2)$ T-duality invariant Killing spinor on $AdS_4\times\mathbb{CP}^3$}\label{sec:SUSYdetailsABJM}
We express the metric of ABJM in terms of the vielbein basis of eq (\ref{eq:frame}). Supersymmetry is preserved in type IIA when the variations of the dilatino and gravitino vanish. For ABJM which has a constant dilaton and zero Romans mass these constraints are
\begin{align}\label{eq: SUSYvar}
\delta \lambda &= \frac{e^{\phi}}{8}\bigg(\frac{3}{2}\slashed{F}_{2}\Gamma^{ab}(i \sigma_2)+\frac{1}{4!} \slashed{F}_{4}(\sigma_1)\bigg)\epsilon =0,\\[2mm]
\delta \Psi_{\mu}&= D_{\mu}\epsilon +\frac{e^{\phi}}{8}\bigg(\frac{1}{2}\slashed{F}_{2}\Gamma^{ab}(i \sigma_2)+\frac{1}{4!} \slashed{F}_{4}( \sigma_1)\bigg)\Gamma_{\mu}\epsilon=0\nn,
\end{align}
where $D_{\mu}\epsilon= (\partial_{\mu}+\frac{1}{4} \omega_{\mu,ab}\Gamma^{ab})\epsilon$. Specifically we have
\beq
\frac{1}{2} \slashed{F}_2= -\frac{2k}{L^2}(\Gamma^{1 6}+ \Gamma^{2 3}+\Gamma^{45}),~~~\frac{1}{4!}\slashed{F}_4= \frac{6k}{L^2}\Gamma_{AdS_4},
\eeq
and
\begin{align}\label{eq:spinconnection}
&\omega^{x^{\mu}\rho}= \frac{2}{L} e^{x^{\mu}},~~~\omega^{12}=-\omega^{36}=\frac{\tan\zeta}{L}e^{2},~~~\omega^{13}=\omega^{26}=\frac{\tan\zeta}{L}e^{3}\nn\\
& \omega^{14}=-\omega^{56}=-\frac{\cot\zeta}{L} e^4,~~~\omega^{15}=\omega^{46}=-\frac{\cot\zeta}{L}e^5,~~~\omega^{16}-\frac{2\cot 2\zeta}{L}e^6,\\
&\omega^{23}=\frac{1}{L} (-2\cot\theta_1\sec\zeta e^3+\tan\zeta e^6),~~~\omega^{45}=\frac{1}{L}(-2\cot\theta_1\sec\zeta e^3+(\cot\zeta+2\tan\zeta )e^6)\nn.
\end{align}

Inserting the fluxes into the variation of the dilatino and manipulating leads to
\beq
\bigg(\Gamma^{23 45}+\Gamma^{1 6}(\Gamma^{23}+\Gamma^{ 45})\bigg)\epsilon= \epsilon.
\eeq
This constraint preserves a maximum of 24 real supercharges, however one finds that such a Killing spinor, which also solves the gravitino variation, must depend on the $SU(2)$ directions \cite{Lozano:2014ata}.
Here we take a different approach and impose the projection
\beq\label{eq: proj1}
\Gamma^{23 45}\epsilon = \epsilon,
\eeq
which preserves only half the supercharges.
Turning attention to the gravitino variation, one finds that the components along the $AdS_4$ directions give
\beq
D_{\mu}\epsilon+\frac{1}{L} \Gamma_{AdS_4}\Gamma_{\mu}(\sigma_1)\epsilon,
\eeq
which is a standard Killing spinor equation for $AdS_4$ which one can solve without any constraint.

Using the projection or eq (\ref{eq: proj1}) it is possible to show that the gravitino variation along the $\mathbb{CP}^3$ directions reduce to a differential equation and an additional projection
\begin{align}
&\partial_{\zeta}\epsilon+ \Gamma^6( i\sigma_2) \epsilon=0,\\[2mm] 
&\Gamma^{1456}\epsilon=-\bigg(\cos2\zeta +\sin 2\zeta \Gamma^6(i \sigma_2)\bigg)\epsilon.
\end{align}
These are solved by
\beq
\epsilon= e^{-\Gamma^6( i\sigma_2)}\epsilon_0
\eeq
where $\epsilon_0$ is a spinor which depends only on the $AdS_4$ coordinates and obeys
\beq
\Gamma^{2345}\epsilon_0=-\Gamma^{1456}\epsilon_0=\epsilon_0.
\eeq
Thus we have found a Killing spinor preserving 8 real supercharges which gives $\mathcal{N}=2$ supersymmetry in 3d. This is the most general spinor which is independent of the $SU(2)$ directions (in the prefered frame) and so \cite{Kelekci:2014ima} informs us that 8 supercharges are preserved under a $SU(2)$ NAT duality transformation.\\

As the solution is a direct product and we know that there are 4 independent Killing spinors preserved by $AdS_4$, we must have  2 preserved on $\mathbb{CP}^3$. On the other hand the $AdS_4$ factor and supersymmety preserved by the spinor imply that we are describing a subsector of ABJM with $U(1)$ R-symmetry. The Killing spinors should be invariant under the action of this $U(1)$. Indeed we can impose an additional projection
\beq
P_{\alpha}\epsilon=\Gamma^{6}(i \sigma_2)\epsilon,~~~~P_{\alpha}=\Gamma^{3}(-\cos\alpha\Gamma^{4}+\sin\alpha\Gamma^5)
\eeq
where $\alpha$ is a constant which parametrizes the $U(1)$. Notice that if one defines two spinors such that $P_{\alpha_1}\chi_{\alpha_1}=\Gamma^{6}(i \sigma_2)\chi_{\alpha_1}$ and $P_{\alpha_2}\chi_{\alpha_2}=\Gamma^{6}(i \sigma_{2})\chi_{\alpha_2}$ hold, then we have $\chi_{\alpha_1}^{\dag}\chi_{\alpha_2}=0$ when $\alpha_1-\alpha_2=\pi$, so we are still describing $\mathcal{N}=2$ supersymmetry. \\

\subsection{A $U(1)$ of $SU(3)$-structures on $\mathbb{CP}^3$}\label{sec:SUSYdetailsABJM2}
We know the 6d Killing spinors of ABJM define an $SU(3)$-structure \cite{Gaiotto:2009yz}, so the internal spinors $\eta^{1}_+$ and $\eta^{2}_+$ must match up to a phase. Specifically we define
\beq
\eta^{1}_+ = e^{i \frac{\theta_0}{2}} \eta_+,~~~\eta^{2}_+ = e^{-i \frac{\theta_0}{2}}\eta_+.
\eeq

The projective constraints in 6d become
\beq
\gamma^{1456}\eta_+= -\bigg(\cos 2\zeta+ \hat{P}_{\alpha} \sin2\zeta \bigg)\eta_+,~~~ \gamma^{2345}\eta_+ =  \eta_+,~~~  \hat{P}_{\alpha}\eta_+ = \gamma^6 \eta_-,
\eeq
where $\hat{P}_{\alpha}=\gamma^{3}(-\cos\alpha\gamma^{4}+\sin\alpha\gamma^5)$. These are still a little complicated, to get to a canonical frame we first rotate in $\gamma^4$, $\gamma^5$, and then $\gamma^3$, $\gamma^4$ such that $\hat{P}_{\alpha}=-\tilde{\gamma}^{34}$ and $\gamma^{1456}\eta_+= -\eta_+$. This leads to new vielbeins which we express in terms of eq (\ref{eq:frame}) as
\begin{align}\label{eq: canframe}
\tilde{e}^a&=e^a,~~~a=1,2,6,\nn\\[2mm]
\tilde{e}^3&=\cos 2 \zeta e^3+ \sin2\zeta(-\cos\alpha e^4+ \sin\alpha e^5),\nn\\[2mm]
\tilde{e}^4&=\sin 2 \zeta e^3+\cos 2 \zeta(\cos\alpha e^4- \sin\alpha e^5),\nn\\[2mm]
\tilde{e}^5&=\sin\alpha e^4 +\cos\alpha e^5.
\end{align}
With respect to this basis we have
\beq
\tilde{\gamma}^{16}\eta_+=\tilde{\gamma}^{32}\eta_+=\tilde{\gamma}^{45}\eta_+= + i \eta_+,~~~\tilde{\gamma}^{346}\eta_+ = -\eta_-,
\eeq
and so the  $SU(3)$-structures are given by the forms
\begin{align}\label{eq: SU3alpha}
J_{\alpha}&= \tilde{e}^{1}\wedge \tilde{e}^{6}+ \tilde{e}^3\wedge \tilde{e}^{2}+ \tilde{e}^4\wedge \tilde{e}^{5},\\[2mm]
\Omega_{hol,\alpha}&= - i ( \tilde{e}^1+ i  \tilde{e}^{6})\wedge (\tilde{e}^3+i \tilde{e}^{2})\wedge (\tilde{e}^4 + i \tilde{e}^5)\nn.
\end{align}
The forms satisfy eq (\ref{eq: diffconSUSY}) for any constant $\alpha$ provided
\beq
\theta_+=\theta_0 = \frac{3\pi}{2},~~~\theta_-=0,~~~~ e^{2A} = \frac{L^2}{4}.
\eeq
One should note that if we take $(J_0,\Omega_{hol,0})$ we can generate the whole $U(1)$ again by sending $\psi \to \psi-\alpha$, inside the left invariant 1-forms $\omega_i$. This is what we expect since the isometry $\partial_{\psi}$  gives the geometric realisation of the $U(1)$ subgroup of the  R-symmetry of ABJM.

\subsection{A $U(1)$ of $SU(2)$-structures in the Non-Abelian T-dual}\label{sec:SUSYdetailsABJMNATD}
We would now like to find the G-structure and Killing spinors of the geometry after performing the $SU(2)$ isometry non-Abelian T-duality. Fortunately we can exploit a map for the $SU(2)$ transformation of the pure spinors that was proposed in \cite{Barranco:2013fza}
\beq\label{eq: puretrans}
\hat \Psi_\pm = \Psi_{\mp}\,\Omega^{-1}_{SU(2)},
\eeq
where in general, in the frame of eq (\ref{eq:frame})
\beq\label{SU2Transformation}
\Omega_{SU(2)}= \frac{1}{\sqrt{1+\zeta_a^2}}\Gamma^{(10)}\bigg(-\Gamma_{456} +\sum_{a=1}^3\zeta_a\Gamma^{a+3}\bigg),
\eeq
for $\zeta_a$ defined as in \cite{Itsios:2013wd}, which for our parametrisation of ABJM specifically is
\beq
\zeta_{1}= \frac{4}{L^2 \cos\zeta\sin^2\zeta}y_1 \cos\xi,~~~\zeta_{2}= \frac{4}{L^2 \cos\zeta\sin^2\zeta}y_1 \sin\xi,~~~ \zeta_3= \frac{4 }{L^2 \sin^2\zeta}y_2.
\eeq
Although eq (\ref{eq: puretrans}) will give us the pure spinors in type IIB, it is still instructive to study the MW Killing spinors. The  action of the NAT duality transformation on this is given by \cite{Kelekci:2014ima} 
\beq
\hat{\epsilon}_1=\epsilon_1,~~~\hat{\epsilon}_2=\Omega_{SU(2)}\, \epsilon_2,
\eeq
which corresponds to the following 6d spinors 
\begin{align}\label{eq: NATDspinors}
\hat\eta^{1}_+ &= e^{i \frac{3\pi}{4}}\eta_+,\\[2mm]
\hat\eta^{2}_+ &= -i\, e^{-i \frac{3\pi}{4}}\bigg[i \,\frac{\cos 2\zeta \tilde{\gamma}^{1} + \tilde{\zeta}_1 \tilde{\gamma}^4 +\tilde{\zeta}_2 \tilde{\gamma}^5+\zeta_3\tilde{\gamma}^{6}}{\sqrt{1+\zeta_a\zeta_a}} \eta_-+\frac{\sin2\zeta}{\sqrt{1+\zeta_a\zeta_a}} \eta_+ \bigg]  \nn,
\end{align}
with the spinors on $AdS_4$ unchanged.
Here we use the frame of eq (\ref{eq: canframe}), but with dual vielbeins, have made use of the projections and defined
\beq
\tilde{\zeta}_{1}= \frac{4}{L^2 \cos\zeta\sin^2\zeta}y_1 \cos(\xi+\alpha),~~~\tilde{\zeta}_{2}= \frac{4}{L^2 \cos\zeta\sin^2\zeta}y_1 \sin(\xi+\alpha).
\eeq
Here we see that $\alpha$ only appears in the combination $\xi+\alpha$, which indicates that $\partial_{\xi}$ plays the role of the  $U(1)$ R-symmetry in the NAT dual solution, indeed this can be confirmed by computing the Kosmann derivative along $\partial_{\xi}$.

The spinors in eq (\ref{eq: NATDspinors}) actually define a dynamical $SU(2)$ structure, which means $\hat\eta^{1}_+$ and $\hat\eta^{2}_+$ are not globally parallel and the angle between them is point dependent. We can simplify the expression for $\hat\eta^{2}_+$ considerably with further rotations of the vielbein basis. There is an optimum frame, in which all components of the $SU(2)$-structure are relatively simple. The vielbeins are given in this frame by
\begin{align}
%
\hat e^{1}&=\frac{1}{4L^3\sin^3\zeta\cos\zeta \sqrt{\Delta_q}}\bigg(L^4\sin^2\zeta\sin4\zeta-32(y_1 dy_1+y_2 dy_2)\bigg),\\[2mm]
\hat e^{2}&=\frac{1}{4L^3\sin^3\zeta\cos\zeta \sqrt{\Delta_p}\sqrt{\Delta_q}}\bigg[\cos\zeta\bigg(\sin 2\zeta\big(L^4\sin^2\zeta\Delta_pd\theta-32y_1^2\sin 2(\xi+\alpha)\sin\theta d\phi\big)\nn\\
&~~~~~~~~~~~~~~~~~~~~~~~~~~~~~~~~~~-128y_1y_2\cos(\xi+\alpha)d\zeta\bigg)-64 y_1\cos(\xi+\alpha)\sin\zeta\cos 2\zeta dy_2  \bigg],\nn\\[2mm]
\hat e^{3}&=\frac{1}{2L^3 \sin^2\zeta\cos\zeta\sqrt{\Delta_0}\sqrt{\Delta_p}}\bigg[\cos^2\zeta \bigg(32y_1y_2\cos(\xi+\alpha)d\xi+32y_2\sin(\xi+\alpha)dy_1+\nn\\
&~~~~~~~~~~~~~~~~\big(32y_1y_2\cos(\xi+\alpha)\cos\theta +L^4\Delta_0\cos2\zeta\sin^2\zeta\sin\theta\big)d\phi \bigg) -32 y_1\sin(\xi+\alpha)d y_2\bigg],\nn\\[2mm]
\hat e^{4}&=\frac{2}{L^5\sin^3\zeta\cos\zeta\sqrt{\Delta_0}\sqrt{\Delta_p}\sqrt{\Delta_q}}\bigg[\cos\zeta\bigg(y_1\sin(\xi+\alpha)\big(64y_1^2\cos^2(\xi+\alpha)+L^4\sin^2\zeta\Delta_p\big)d\xi\nn\\
& ~~~~~~~~~~~~~~~~~+\cos(\xi+\alpha)\big(64y_1^2\sin^2(\xi+\alpha)dy_1 + 64 y_1y_2dy_2 - L^4\Delta_p\sin^2\zeta dy_1\big)\phantom{\bigg.}  \nn\\
&~~~~~~~~~~~~~~~~~+L^4y_1\Delta_q\sin(\xi+\alpha)\sin^2\zeta\cos\theta d\phi\bigg)-2L^4y_1\Delta_0 \cos(\xi+\alpha)\cos2\zeta\sin\zeta d\zeta\bigg],\nn\\[2mm]
\hat e^{5}&=-\frac{2}{L\sin^2\zeta \cos\zeta\sqrt{\Delta}_p}\bigg[2y_1\sin(\xi+\alpha)d\zeta-y_2\sin2\zeta\cos\zeta^2\sin\theta d\phi+\nn\\
&~~~~~~~~~~~~~~~~~~~~~~~~~~~~~~~~~~~~~~~\frac{1}{4}\sin4\zeta\bigg(\sin(\xi+\alpha)dy_1+y_1\cos(\xi+\alpha)\big(d\xi+\cos\theta d\phi\big)\bigg)\nn\\[2mm]
\hat e^{6}&=-\frac{2}{L\sin^2\zeta\cos\zeta\sqrt{\Delta_q}}\bigg[\cos2\zeta\sin\zeta dy_2+\nn\\
&~~~~~~~~~~~~~~~~~~~~~~~~~~~~2\cos\zeta\bigg(y_2d\zeta+y_1 \cos\zeta \sin\zeta\big(\cos(\xi+\alpha)d\theta+\sin(\xi+\alpha)\sin\theta d\phi\big)\bigg)\bigg]\nn,
\end{align}
where
\beq
\Delta_0=1 +\zeta_a^2,~~~\Delta_q =\cos^2 2\zeta +\zeta_a^2,~~~ \Delta_p= \Delta_q-\sin^2 2\tilde{\zeta} \zeta_1^2.
\eeq
In this basis the action of NAT duality on the 6d spinors is simply
\beq
\hat\eta^1=e^{i \frac{\theta_0}{2}}\eta_+,~~~~~\hat \eta^2_+ = -e^{-i \frac{\theta_0}{2}}\big[\kappa_{\|} \eta_+ + i \kappa_{\perp}\hat\gamma^1\eta_-\big]
\eeq
where
\beq\label{eq: kappas}
\kappa_{\|}=\frac{\sin2\zeta}{\sqrt{1+\zeta_a\zeta_a}},~~~~\kappa_{\perp}=\sqrt{\frac{\cos2\zeta+\zeta_a\zeta_a}{1+\zeta_a\zeta_a}}
\eeq
and $\kappa_{\|}^2+\kappa_{\perp}^2=1$. The projections the original spinor obeys are most succinctly expressed as
\beq
\hat\gamma_{2345}\eta_+=\eta_+,~~~\hat\gamma_{1456}\eta_+ = -(\kappa_{\perp}- \kappa_{\|} \hat\gamma^{34})\eta_+,
\eeq
in the basis where $\gamma^{(7)}\eta_+=\eta_+$ as before.
The $U(1)$'s worth of $SU(2)$-structure is given by the following forms
\begin{align}
z_{\alpha}&=\hat e^1+i \hat e^6,\nn\\[2mm]
j_{\alpha}&=   (\kappa_{\perp}\hat e^3-\kappa_{\|}\hat e^4)\wedge \hat e^2+ (\kappa_{\perp}\hat e^4 +\kappa_{\|}\hat e^3)\wedge \hat e^5,\nn\\[2mm]
\omega_{hol,\alpha}&=-i\big((\kappa_{\perp} \hat e^3- \kappa_{\|} \hat e^4)+ i\hat e^2\big)\wedge\big((\kappa_{\perp} \hat e^4+\kappa_{\|} \hat e^3)+i\hat e^5\big),
\end{align}
which satisfy the supersymmetry conditions of eq (\ref{eq: diffconSUSY}) for any constant $\alpha$ provided
\beq
\theta_+=0,~~~\theta_-=\theta_0 = \frac{3\pi}{2},~~~~ e^{2A} = \frac{L^2}{4}.
\eeq
We could take the intersection of the two linearly independent $SU(2)$ structures defined for $\alpha=0$ and $\alpha=\pi$, and define an identity structure. However, the supersymmetry conditions of such an object are absent from the literature at present and deriving them is outside the scope of this work.

\subsection{Killing spinors in M-theory}\label{sec:Mtheoryspinors}
Before we can define the M-theory Killing spinor, we must first derive the MW Killing spinors in type IIA after an additional T-duality is performed. As we want to make contact with \cite{Gabella:2012rc} we need to work with the two linearly independent spinors in 6d. These are given by
\beq
\pi^{1}_+ = e^{i \frac{\theta_0}{2}}\pi_+,~~~\pi^{2}_+ = e^{-i \frac{\theta_0}{2}}\pi_+,~~~\tilde{\pi}^{1}_+ = e^{i \frac{\theta_0}{2}} \tilde{\pi}_+,~~~\tilde{\pi}^{2}_+ = e^{-i \frac{\theta_0}{2}}\tilde{\pi}_+,~~~~\theta_0 =\frac{3\pi}{2}
\eeq
and are such that
\begin{align}\label{eq:piprojections}
&\gamma^{1456}\pi_+= -\big(\cos2\zeta-\sin2\zeta \gamma^{34}\big)\pi_+,~~~\gamma^{2345}\pi_+=\pi_+,~~~\gamma^{246}\pi_+=-\pi_-\nn\\[2mm]
&\gamma^{1456}\tilde\pi_+= -\big(\cos2\zeta+\sin2\zeta \gamma^{34}\big)\tilde\pi_+,~~~\gamma^{2345}\tilde\pi_+=\tilde\pi_+,~~~\gamma^{246}\tilde\pi_+=\tilde\pi_-,
\end{align}
in the frame of eq (\ref{eq:frame}). The independent 10d MW spinors $\epsilon_{1,2}$ and $\tilde{\epsilon}_{1,2}$ are then constructed  in the obvious way from eq (\ref{eq: MWSpinors}), with $\eta\to\pi$ and using the same spinors on $AdS_4$.  We must act on these spinors first with $\Omega_{SU(2)}$, which in this frame is as in  eq (\ref{SU2Transformation}), then with $\Omega_{U(1)}$, which gives the transformation of the spinor under the Abelian T-duality \cite{Hassan:1999bv}. In the frame of eq (\ref{eq:frame},\ref{eq:frame2}) this is most succinctly expressed as
\beq
\Omega_{U(1)}=\frac{1}{\sqrt{\Delta_0}\sqrt{\Xi_0}}\sin \zeta\cos\theta\Gamma^{(10)}\bigg[\big(-\zeta_2+\zeta_1\zeta_3\big)\Gamma^4+\big(\zeta_1+\zeta_2\zeta_3\big)\Gamma^5-\big(\zeta_1^2+\zeta_2^2\big)\Gamma^6\bigg]-\frac{\sqrt{\Delta_0}}{\sqrt{\Xi_0}}\sin\theta\Gamma^{(10)}\Gamma^3\nn.
\eeq 
We take the 10d MW Killing spinors in IIA after the NAT-T duality transformation to be
\beq
\hat{\hat{\epsilon}}_1 =\epsilon_1,~~~\hat{\hat{\epsilon}}_2= \Omega_{U(1)}\Omega_{SU(2)}\epsilon_2,
\eeq
with an equivalent expression with $\epsilon\to\tilde{\epsilon}$, which means that the new 6d Killing spinors are given by
\begin{align}\label{eq:IIAspinors}
\hat{\hat{\pi}}^1_+=& e^{i \frac{\theta_0}{2}}\pi_+,~~~~~~
\hat{\hat{\pi}}^2_+=-e^{-i \frac{\theta_0}{2}}\bigg(\hat{\kappa}_{\|}\pi_++\hat{\kappa}_{\perp}\mathcal{F}\pi_- \bigg),\nn\\[2mm]
\hat{\hat{\tilde{\pi}}}^1_+=& e^{i \frac{\theta_0}{2}}\tilde{\pi}_+,~~~~~~
\hat{\hat{\tilde{\pi}}}^2_+= + e^{-i \frac{\theta_0}{2}}\bigg(\hat{\kappa}_{\|}\tilde{\pi}_++\hat{\kappa}_{\perp}\mathcal{F}\tilde{\pi}_-\bigg),
\end{align}
where
\begin{align}
\hat{\kappa}_{\|}&=\frac{\sin2\zeta\sin\theta\zeta_2}{\sqrt{\Xi_0}},~~~\hat{\kappa}_{\perp}=\sqrt{1-\hat{\kappa}^2_{\|}},\nn\\[2mm]
 \mathcal{F}&=\frac{i}{\sqrt{\Xi_0 -\sin^2 2\zeta \sin^2\theta \,\zeta_2^2}}\bigg(\sin\zeta\cos\theta\big(\zeta_2\gamma^2-\zeta_1\gamma^3\big)-\sin\theta\big(\cos2\zeta\,\zeta_2\gamma^1-\gamma^5-\zeta_3\gamma^4+\zeta_1\gamma^6\big)\bigg).
\end{align}
Clearly eq (\ref{eq:IIAspinors}) supports a $U(1)$ of dynamical $SU(2)$-structures, as was the case in type-IIB, which we will not explicitly derive.\\

We are know ready to construct the two independent M-theory Killing spinors. These can be expressed in terms of the spinors in IIA as
\beq
\eta^1 = e^{-\Phi/6}\big(\epsilon_1+\epsilon_2),~~~ \eta^2=e^{-\Phi/6}\big(\tilde{\epsilon}_1+\tilde{\epsilon}_2).
\eeq
In the conventions of \cite{Gabella:2012rc} the M-theory spinors are
\beq\label{eq:sdg}
\eta^{i}= e^{\tilde{\Delta}/2}\bigg( \psi^i_+\otimes \chi_i+ \big(\psi^i_+\big)^c\otimes \chi^c_i\bigg),
\eeq
where $ e^{\frac{\tilde{\Delta}}{2}}=e^{\frac{\tilde{A}}{2}-\frac{\Phi}{6}}$ and $e^{2\tilde{A}}$ is a modified  warp factor of $AdS_4$ in IIA such that $Ricci(AdS_4)=-12 g(AdS_4)$. Thus if we identify the $AdS_4$ spinors of IIA with those of eq (\ref{eq:sdg}) we see that
\begin{align}
\chi_1&=\frac{1}{\sqrt{2}}\big(\hat{\hat{\pi}}^1_++\hat{\hat{\pi}}^2_-\big),~~~~~~
\chi^c_1=\frac{1}{\sqrt{2}}\big(\hat{\hat{\pi}}^1_-+\hat{\hat{\pi}}^2_+\big),\nn\\[2mm]
\chi_2&=\frac{1}{\sqrt{2}}\big(\hat{\hat{\tilde{\pi}}}^1_++\hat{\hat{\tilde{\pi}}}^2_-\big),~~~~~~~
\chi^c_2=\frac{1}{\sqrt{2}}\big(\hat{\hat{\tilde{\pi}}}^1_-+\hat{\hat{\tilde{\pi}}}^2_+\big),
\end{align}
which clearly satisfy $\bar{\chi}_1\chi_1=\bar{\chi}_2\chi_2=1$, and from these one can construct spinors of charge $\pm 2$ under  the $U(1)$ R-symmetry
\beq
\chi_{\pm} =\frac{1}{\sqrt{2}}\big(\chi_1\pm\chi_2\big).
\eeq
It is then simply a matter of plugging the $\chi_{\pm}$ of this section into the spinor bi-linears in appendix B of \cite{Gabella:2012rc}, and rotating the frame to reproduce the results of section (\ref{sec: SU2}). Note that the frame used in this section needs to be rotated as in eq (\ref{eq:framerotation}) to reach the vielbein basis  where flat directions 2456 may be identified with $\mathcal{G}^{1,2,3,4}$ of eq (\ref{eq: Gs}) and the rest with eq (\ref{eq:frame}).

\end{document}